\documentclass[12pt]{article}
\usepackage[top=1.in, bottom=1in, left=1.1in, right=1.1in]{geometry}

\usepackage{color}
\usepackage{graphicx}
\usepackage[small]{caption}
\usepackage{authblk}
\usepackage{bm}
\usepackage{booktabs}
\usepackage{newtxtext}
\usepackage{newtxmath}
\usepackage[round]{natbib}
\usepackage{hyperref}
\hypersetup{
    colorlinks = true,
    urlcolor   = blue,
    citecolor  = blue,
    linkcolor  = blue,
}
\usepackage{enumitem}
\usepackage{multirow}

\newcommand\Real{\mbox{Re}}          
\newcommand\Rey{\mbox{\textit{Re}}}  

\title{Reduced-order variational mode decomposition}

\author[1]{Zi-Mo Liao}
\author[1]{Zhiye Zhao}
\author[1]{Liang-Bing Chen}
\author[1]{Zhen-Hua Wan\thanks{Email address for correspondence: wanzh@ustc.edu.cn}}
\author[1]{Nan-Sheng Liu\thanks{Email address for correspondence: lns@ustc.edu.cn}}
\author[1]{Xi-Yun Lu}

\affil[1]{Department of Modern Mechanics, University of Science and Technology of China, Hefei 230027, PR China}

\begin{document}
\maketitle

\begin{abstract}
	A novel data-driven method of modal analysis for complex flow dynamics, termed as reduced-order variational mode decomposition (RVMD), has been proposed, combining the idea of the separation of variables and a state-of-the-art nonstationary signal-processing technique -- variational mode decomposition. It enables a low-redundant adaptive extraction of coherent structures in statistically nonstationary flows, with its modes computed by solving an elaborate optimization problem using the block coordinate descent algorithm. Discussion on the intrinsic relations between RVMD and some classic modal decomposition methods demonstrates that RVMD can be reduced into proper orthogonal decomposition (POD) or discrete Fourier transform (DFT) at particular parameter settings. The significant advantages of RVMD for performing time-frequency analysis are highlighted by a signal-processing analogous categorization of the widely-used modal decomposition techniques. It is also confirmed that the combination of RVMD and the Hilbert spectral analysis provides a physically intuitive way to explore the space-time-frequency characteristics of transient dynamics. Finally, all the appealing features of RVMD mentioned above are well verified via two canonical flow problems: the transient cylinder wake and the rectangular turbulent supersonic screeching jet.
\end{abstract}

\newpage
\tableofcontents

\section{Introduction}
\label{sec:intro}
Modal decomposition has become fundamentally important in turbulence research in terms of constructing a low-dimensional representation in the Eulerian perspective and establishing an explicit connection of dynamic process between the physical and phase space \citep{Holmes12}. Since \cite{Lumley67,Lumley70} proposed the proper orthogonal decomposition (POD) in the realm of turbulence, a large variety of modal decomposition methods have been developed for more than 50 years. Coherent structures -- organized fluid elements of significant lifetime and scale -- can be extracted by these well-designed methods, enabling the mechanistic study of the essential dynamical features inherent in complex flows. Modal analysis has demonstrated its powerful roles in theoretical analysis and engineering applications \citep{Taira17,Taira20}, for example, shedding light on the potential mechanisms of the unsteadiness in shock wave/boundary layer interactions \citep{Priebe16}, self-similar behaviour in pipe flows \citep{Hellstrom16} and providing insight into the dynamics of large-scale wavepackets in turbulent jets \citep{Schmidt18}, among others.

Some general consensus has been reached on the direction of the development of modal decomposition, even though various techniques are rooted in different physical considerations and mathematical operations. As proposed by \citet{Noack16}, the first is to extend the information contained in the decomposed modes, and the second is to extract a sparse description of the dominant feature in the original dynamic system \citep{Brunton16}. Although increasingly sophisticated experimental techniques and high-performance computing have provided a huge mass of time-accurate flow field data, namely, a prerequisite to explore the transient and intermittent behaviours in statistically nonstationary flows, further development of a mode-based time-frequency analysis framework is still highly desired~\citep{Nekkanti21}. It is due to the fact that the conventional construction ideas of existing methods, such as time averaging and linear approximation, limit their ability to characterize the time-frequency information, which motivates us to introduce a new mathematical form of modal expansion to overcome this difficulty.

The two most popular modal analysis techniques, POD and dynamic mode decomposition \citep[DMD,][]{Schmid10}, have demonstrated their good capabilities in a wide range of problems, although various attempts are still raised for their more extensive applications. The core idea of POD is to find an orthonormal basis that spans a finite-dimensional subspace, such that the $L^2$-norm of the projected data onto this subspace obtains its maximum in a time (or ensemble) averaged sense \citep{Holmes12}. This energetic optimality makes it an efficient way to compress the original data, but the averaging process may result in the loss of the key dynamical information. DMD is proposed to extract modes with temporal monochromaticity through a linear approximation of the original nonlinear system, which can also be regarded as a finite-dimensional approximation of the Koopman operator \citep{Rowley09,Brunton22}. The eigenvalues and corresponding eigenvectors of the approximated system comprise the DMD modes, whose time coefficients are pure harmonics with exponential growing or decaying. For mean-subtracted data, \citet{Chen12} have proved that DMD is equivalent to the discrete Fourier transform (DFT). Different from POD, DMD allows us to examine the spectral content of the system in a global manner. Nevertheless, the exponential oscillatory formulation of the time coefficients limits its application to describe general nonstationary processes.

Recently, various time-frequency analysis strategies have been attempted by incorporating concepts in signal processing into the existing modal decomposition methods. \citet{Kutz16} presented a recursive algorithm called multiresolution DMD, in which the modes with the frequency characteristics localized in time are obtained by a removal-splitting operation. Another interesting attempt was made by \citet{Nekkanti21} based on the spectral proper orthogonal decomposition \citep[SPOD,][]{Towne18}. SPOD is a space-time formulation of POD, which inherits the basic idea of \citet{Lumley67,Lumley70}. For statistically stationary flows, it computes orthogonal modes that diagonalize the estimated cross-spectral density matrix at each frequency. Two approaches were presented in \citet{Nekkanti21} to recover the time-frequency contents from the already computed SPOD modes: oblique projection in the time domain and convolution in the frequency domain. These frameworks mentioned above can serve as a good pilot to guide the development of mode-based time-frequency analysis while further efforts are required to improve performance and operability.

In this study, we develop a new data-driven modal analysis method referred to as reduced-order variational mode decomposition (RVMD), which is inspired by the Hilbert-Huang transform \citep[HHT,][]{Huang98} and a state-of-the-art signal-processing technique called variational mode decomposition \citep[VMD,][]{Dragomiretskiy14}. It is well known that HHT can provide a feasible path for dealing with nonstationary, nonlinear signals in the Hilbert view \citep{Huang99}, which is different from the Fourier-based methods such as short-time Fourier transform and wavelet transform. HHT is implemented by decomposing the original time series into intrinsic mode functions (IMFs) and then applying Hilbert spectral analysis to obtain the instantaneous frequencies. As the core concept of HHT, IMFs are usually obtained by empirical mode decomposition or VMD. IMF represents a generalized Fourier expansion that can effectively characterize time-frequency contents with superior resolution~\citep{Huang99}. Drawing on the construction of VMD and combining the idea of the separation of variables, the proposed RVMD can adaptively extract time-frequency features from spatio-temporal data and obtains excellent physical interpretability.
The adaptive frequency-domain sparsification is achieved by an iterative optimization using the block coordinate descent algorithm \citep{Wright15}, similar to the idea of sparsity-promoting DMD \citep{Jovanovic14}. Note that, the appealing features of RVMD are in line with the outlook in \citet{Noack16}, as the direct results of a mathematically well-defined optimization problem. Two canonical flow problems, i.e., the transient cylinder wake and the rectangular turbulent supersonic screeching jet, are employed to validate the capability of RVMD. It is demonstrated that RVMD not only captures the dominant temporal dynamics accurately but also provides physically meaningful modal information.

The remainder of this paper is organized as follows. Section \ref{sec:problem} formulates the proposed method and discusses the properties of computed modes, as well as the criteria for parameters setting. Section \ref{sec:relation} elucidates the intrinsic relations between RVMD and various existing methods based on a mathematical perspective, and then presents a signal-processing analogous categorization of different modal techniques. In \S~\ref{sec:application}, two canonical cases are utilized to demonstrate the capabilities and advantages of RVMD. Finally, concluding remarks are summarized in \S~\ref{sec:conclusion}.

\section{Reduced-order variational mode decomposition}
\label{sec:problem}
\subsection{Problem formulation}
Similar to existing modal decomposition methods, RVMD attempts to find a low-rank variables-separation representation of the observed space-time data $q(\boldsymbol{x},t)$ as the following form
\begin{equation}
	q(\boldsymbol{x},t)=\sum_k\phi_k(\boldsymbol{x})c_k(t),
\end{equation}
where $\phi_k(\boldsymbol{x}),\ \boldsymbol{x}\in\Omega$ denotes the spatial distribution of the $k$-th mode and $c_k(t),\ t\in(-\infty,\infty)$ is the time-evolution coefficient, with $\Omega$ denoting the spatial domain over which the flow is defined. Due to the finite energy nature of the flow field, all these functions are square-integrable. Following the formulation of POD \citep{Holmes12}, we further assume that they belong to the Hilbert space with inner product $\langle\cdot,\cdot\rangle$ and $L^2$-norm $\|\cdot\|$ \citep{Zeidler12}. Specific definitions of the norm in space and time are
\begin{eqnarray}
	\|\phi(\boldsymbol{x})\|_{\boldsymbol{x}}&\equiv&\langle\phi,\phi\rangle^{1/2}_{\boldsymbol{x}}\quad\text{with}\quad\left\langle \phi_1(\boldsymbol{x}),\phi_2(\boldsymbol{x})\right\rangle_{\boldsymbol{x}}=\int_\Omega \phi_1(\boldsymbol{x})\overline{\phi_2(\boldsymbol{x})}\mathrm{d}\boldsymbol{x},
\end{eqnarray}
and
\begin{eqnarray}
	\|c(t)\|_t&\equiv&\langle c,c\rangle^{1/2}_t\quad\text{with}\quad\left\langle c_1(t),c_2(t)\right\rangle_t=\int_{-\infty}^\infty c_1(t)\overline{c_2(t)}\mathrm{d}t,
\end{eqnarray}
respectively. The overline denotes complex conjugate.

The proposed RVMD modes are a group of triplets with a finite number $K$, written as
\begin{eqnarray}
	&&\left.\left\{\phi_k(\boldsymbol{x}),c_k(t),\omega_k\right\}\right|_{k=1}^K,
	\label{eqn:rvmdmodes}
\end{eqnarray}
where $\omega_k$ is called the central frequency of $c_k(t)$ that, as shown below, is the averaged frequency weighted by the square of the Fourier transform of the time coefficient $|\hat{c}_k(\omega)|^2$, with $\ \hat{c}_k(\omega)=\mathcal{F}\{c_k(t)\}=\int_{-\infty}^\infty c_k(t)e^{-\mathrm{i}\omega t}\mathrm{d}t$ and $\,\hat{\cdot}\,$ denoting the Fourier transform hereafter. Domains of $\phi_k(\boldsymbol{x})$, $c_k(t)$, and $\omega_k$ are
\begin{equation}
	\Phi\equiv\left\{\phi(\boldsymbol{x})\in L^{2,\mathbb{R}}(\Omega)\big|\|\phi(\boldsymbol{x})\|_{\boldsymbol{x}}=1\right\},\quad C\equiv L^{2,\mathbb{R}}(-\infty,\infty),\quad \mathbb{R}^+,
	\label{eqn:region1}
\end{equation}
respectively. Note that, all the components of the RVMD modes are defined in the real domain, ensuring a straightforward flow field reconstruction and a clear physical meaning. Besides, since the spatial distributions are normalized, the energy of each mode can be reflected by its time coefficient, i.e.,
\begin{equation}
	E_k=\|c_k(t)\|_t^2.
\end{equation}
The energy ratio can be further defined to measure the relative strength of each mode
\begin{equation}
	\tilde{E}_k\equiv \frac{E_k}{\left(\sum_{i=1}^K{E_i}\right)}.
\end{equation}

\begin{figure}
	\centerline{
		\includegraphics[width=0.98\linewidth]{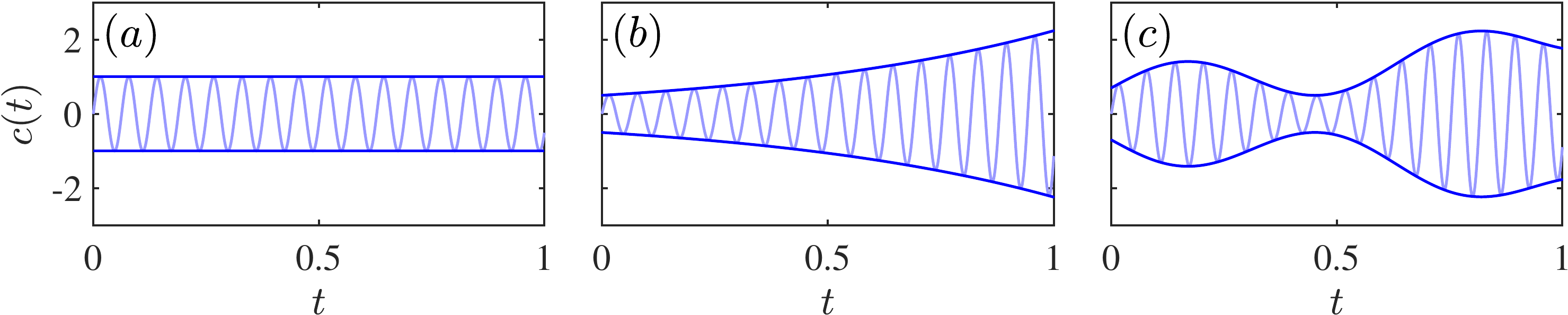}}
	\caption{Temporal signals $c(t)$: ($a$) pure harmonic, ($b$) pure harmonic with exponential growing, ($c$) an example of intrinsic mode function. Each signal's upper/lower envelope is indicated using a dark blue curve.}
	\label{fig:imf}
\end{figure}

Before we present the optimization problem that RVMD deals with, it is necessary to introduce the so-called IMF. Since \citet{Huang98} first proposed the idea of IMF, it has become the central concept of modern nonstationary signal-processing techniques. IMFs are amplitude-modulated-frequency-modulated signals written as:
\begin{equation}
	c(t)=A(t)\cos\varphi(t),
	\label{eqn:imf}
\end{equation}
with non-decreasing phase $\varphi(t)$ and non-negative envelope $A(t)$. Besides, the envelope and the instantaneous frequency $\mathrm{d}\varphi/\mathrm{d}t$ should vary much slower than the phase \citep{Gilles13,Daubechies11}. IMFs can be seen as an extension of the Fourier expansion, while the slow-varying envelope contains information on the temporal evolution (see figure \ref{fig:imf}). The above definition yields an immediate consequence of limited bandwidth. On the other hand, if an oscillatory signal has a compact bandwidth around a specific frequency in the spectral domain, it will act as an IMF in the time domain. The concept of IMFs enables RVMD to have the expected advantage in describing time-frequency characteristics. In addition, the formulation of RVMD relies on some other fundamental definitions and operations in signal processing, which are given in appendix \ref{app:fundamentals}.

The core idea of RVMD is to find $K$ modes, as indicated in (\ref{eqn:rvmdmodes}), whose time coefficients are IMFs, providing a low-redundant approximation to the observed field $q(\boldsymbol{x},t)$. We achieve this goal by limiting the time coefficient to have a compact bandwidth, rather than explicitly introducing the formulation of IMF (\ref{eqn:imf}). In other words, we compute the RVMD modes by minimizing the following two components: the measure of the deviation between the original flow and the mode-reconstructed one, and the sum of the bandwidths of the time-evolution coefficients. The former is expressed as a quadratic penalty term, and the latter is estimated through the squared $L^2$-norm of the gradient of $c_k(t)$, as treated in VMD \citep{Dragomiretskiy14}. The resulting unconstrained optimization problem is given by
\begin{eqnarray}
	\min_{\left.\left\{\phi_k(\boldsymbol{x}),c_k(t),\omega_k\right\}\right|_{k=1}^K}&&
	\Bigg\{
	\alpha\sum_{k=1}^K\bigg\|\partial_t\left\{\left[\left(\delta(t)+\mathrm{i}\frac{1}{\pi t}\right)*c_k(t)\right]e^{-\mathrm{i}\omega_k t}\right\}\bigg\|_t^2 \notag\\
	&&\quad+\bigg\|q(\boldsymbol{x},t)-\sum_{k=1}^K\phi_k(\boldsymbol{x})c_k(t)\bigg\|_\mathrm{F}^2
	\Bigg\},
	\label{eqn:rvmd}
\end{eqnarray}
where $\delta(\cdot)$ is the Dirac delta function, $*$ denotes convolution in the time domain. The convolution of $\delta(t)+\mathrm{i}/\pi t$ and $c_k(t)$ leads to the analytic representation of $c_k(t)$. An exponential with frequency $\omega_k$ is multiplied to shift the mode's frequency spectrum to the baseband, using the frequency shifting property of analytic signals (\ref{app:analytic}).
We will see later that this manipulation results in a Wiener filtering (\ref{sec:wiener}) around the central frequency $\omega_k$.
The norm $\|f\|_\mathrm{F}$ for a given space-time function $f(\boldsymbol{x},t)$ in $L^{2,\mathbb{R}}(\Omega)\times L^{2,\mathbb{R}}(-\infty,\infty)$ is defined as
\begin{equation}
	\|f(\boldsymbol{x},t)\|_\mathrm{F}^2=\int_\Omega\int_{-\infty}^\infty f(\boldsymbol{x},t)\overline{f(\boldsymbol{x},t)}\ \mathrm{d}t\mathrm{d}\boldsymbol{x},
\end{equation}
which corresponds to the Frobenius norm of matrices. The positive regularization parameter $\alpha$ in (\ref{eqn:rvmd}), referred to as the filtering parameter hereafter, reflects the relative importance of compact bandwidth and accurate reconstruction. The two parameters, $\alpha$ and $K$, should be input as a priori.

For mathematical simplicity, a spectral-domain alternative of (\ref{eqn:rvmd}) is utilized to compute the RVMD modes. Since the observed data and the RVMD modes are real-valued square-integrable functions, Parseval's theorem can be applied. For two functions $f(t),g(t)\in L^{2,\mathbb{C}}(-\infty,\infty)$, this theorem gives the following identity
\begin{equation}
	\int_{-\infty}^\infty f(t)\overline{g(t)}\mathrm{d}t=\frac{1}{2\pi}\int_{-\infty}^\infty \hat{f}(\omega)\overline{\hat{g}(\omega)}\mathrm{d}\omega.
	\label{eqn:parseval1}
\end{equation}
When $g(t)=f(t)$, (\ref{eqn:parseval1}) turns into
\begin{equation}
	\|f(t)\|_t^2=\frac{1}{2\pi}\|\hat{f}(\omega)\|_\omega^2\quad\text{with}\quad\|\hat{f}(\omega)\|_\omega^2=\int_{-\infty}^\infty |\hat{f}(\omega)|^2\mathrm{d}\omega.
	\label{eqn:parseval2}
\end{equation}
Then, the first term in (\ref{eqn:rvmd}) can be converted into
\begin{equation}
	\frac{1}{2\pi}\alpha\sum_{k=1}^K\bigg\|\mathrm{i}\omega[1+\mathrm{sgn}(\omega+\omega_k)]\hat{c}_k(\omega+\omega_k)\bigg\|_\omega^2,
	\label{eqn:norm1}
\end{equation}
where $\hat{c}_k(\omega)$ is defined in
\begin{equation}
	\hat{C}=\left\{\hat{c}(\omega)\in L^{2,\mathbb{C}}(-\infty,\infty)\big|\hat{c}(-\omega)=\overline{\hat{c}(\omega)}\right\}.
	\label{eqn:region2}
\end{equation}
Performing a variable substitution $\omega\leftarrow\omega-\omega_k$, the norm (\ref{eqn:norm1}) can be written as two times the integral over the non-negative frequencies
\begin{equation}
	\frac{1}{2\pi}\sum_{k=1}^K\int_{0}^\infty 4\alpha (\omega-\omega_k)^2 |\hat{c}_k(\omega)|^2\mathrm{d}\omega.
	\label{eqn:bandwidth}
\end{equation}
Similarly, considering the Hermitian symmetry of real-valued functions, the quadratic penalty term becomes
\begin{equation}
	\frac{1}{2\pi}\int_\Omega\int_{0}^\infty 2\bigg|\hat{q}(\boldsymbol{x},\omega)-\sum_{k=1}^K\phi_k(\boldsymbol{x})\hat{c}_k(\omega)\bigg|^2\mathrm{d}\omega\mathrm{d}\boldsymbol{x}.
	\label{eqn:penalty}
\end{equation}

Finally, we get the following optimization problem in the spectral domain, which is equivalent to (\ref{eqn:rvmd}),
\begin{equation}
	\min_{\left.\left\{\phi_k(\boldsymbol{x}),\hat{c}_k(\omega),\omega_k\right\}\right|_{k=1}^K} F\left(\left.\left\{\phi_k(\boldsymbol{x}),\hat{c}_k(\omega),\omega_k\right\}\right|_{k=1}^K\right),
	\label{eqn:rvmdspectral}
\end{equation}
where the objective function $F\left(\cdot\right)$ is
\begin{eqnarray}
	&&F\left(\left.\left\{\phi_k(\boldsymbol{x}),\hat{c}_k(\omega),\omega_k\right\}\right|_{k=1}^K\right)\equiv\notag\\
	&&\qquad\sum_{k=1}^K\int_{0}^\infty 2\alpha (\omega-\omega_k)^2 |\hat{c}_k(\omega)|^2\mathrm{d}\omega \notag\\
	&&\qquad+\int_\Omega\int_{0}^\infty \bigg|\hat{q}(\boldsymbol{x},\omega)-\sum_{k=1}^K\phi_k(\boldsymbol{x})\hat{c}_k(\omega)\bigg|^2\mathrm{d}\omega\mathrm{d}\boldsymbol{x}.
\end{eqnarray}

The objective function is to be minimized over all the three components of the RVMD modes in the corresponding feasible regions: $\phi_k(\boldsymbol{x})\in\Phi$, $\hat{c}_k(\omega)\in\hat{C}$, and $\omega_k\in\mathbb{R}^+$. The solving process is operated over the positive half of the frequency domain, making use of the Hermitian symmetry, and then the time coefficients $c_k(t)$ can be reconstructed directly. The adaptivity is achieved by optimizing over the central frequencies, thanks to the good mathematical properties of the analytic representation of IMFs. For appropriate parameters setup, RVMD can find modes oscillating around specific frequencies with limited bandwidths, which have (locally) energetic dominance compared to the surrounding broadband noise due to turbulent motions and reflect distinctive dynamical properties coherent in space and time.

Moreover, the spatial distribution functions $\phi_k(\boldsymbol{x})$ are not restricted to be orthogonal in space. Instead, a characteristic scale separation is realized in the time domain \citep{Huang99}. That is, RVMD inherits the advantages of DMD that the non-orthogonal modes provide the ability to capture the essential dynamical behaviours in systems with non-normal dynamical generators \citep{Trefethen93,Schmid07,Jovanovic14}.

\subsection{Computing the RVMD modes}
\begin{figure}
	\centerline{
		\includegraphics[width=0.48\linewidth]{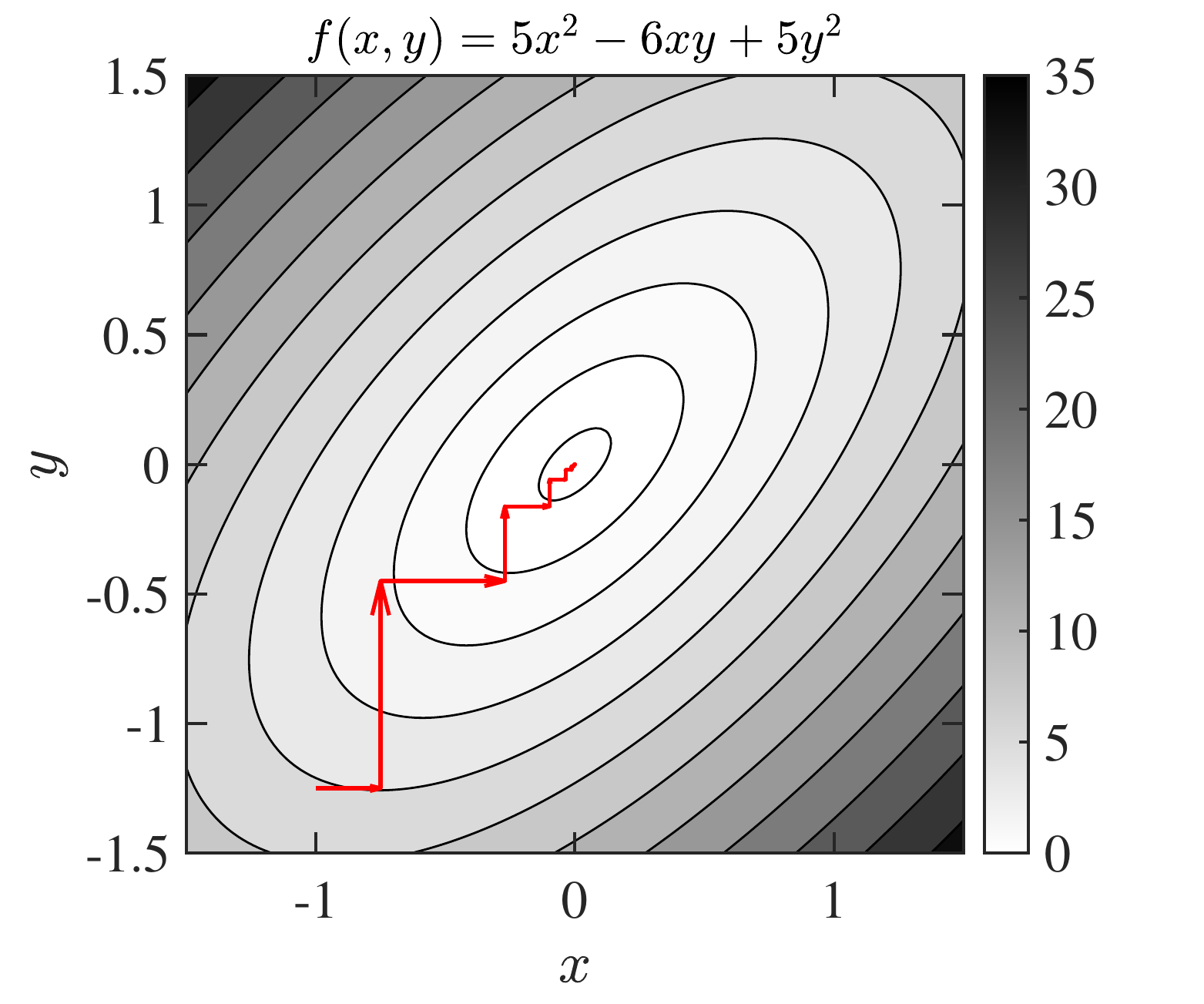}
	}
	\caption{The iteration trajectory of an example problem solved by the coordinate descent algorithm: minimize $f(x,y)=5x^2-6xy+5y^2$ over the two variables $x$ and $y$ for a given initial value $(x^0,y^0)=(-1,-1.25)$. Sub-minimizations on different coordinates are solved successively: $x^{n+1}=\operatorname{argmin}_x\{f(x,y^n)\}$; $y^{n+1}=\operatorname{argmin}_y\{f(x^{n+1},y)\}$. As the iteration progresses, $(x^n,y^n)$ approaches the local minimum, which is also the global minimum as $f$ is convex.
	}
	\label{fig:bcd}
\end{figure}

Although RVMD adopts an optimization-problem-based framework similar to POD, the former is relatively more complicated to solve. In POD and SPOD, the constructed optimization can be directly converted into an equivalent eigenvalue problem that is easy to solve \citep{Holmes12,Towne18}. To deal with the problem (\ref{eqn:rvmdspectral}), we employ the block coordinate descent (BCD) algorithm described by \cite{Liu20}. The BCD-type algorithms solve a global optimization by successively performing approximate minimization along coordinate hyperplanes. Since the BCD-type algorithms are of efficient performance and easy to implement for handling large-scale non-convex optimization problems, they have been widely applied in computational statistics and machine learning \citep{Wright15}. An illustrative example of the coordinate descent algorithm is depicted in figure \ref{fig:bcd}.

To compute the RVMD modes, we first define auxiliary functions as follows
\begin{eqnarray}
	F_{\phi_k}^{n+1}(\phi_k)
	&\equiv& F\left(\phi_1^{n+1},\hat{c}_1^{n+1},\omega_1^{n+1};\dots;\phi_k,\hat{c}_k^{n},\omega_k^{n};\dots;\phi_K^{n},\hat{c}_K^{n},\omega_K^{n}\right), \\
	F_{\hat{c}_k}^{n+1}(\hat{c}_k)
	&\equiv& F\left(\phi_1^{n+1},\hat{c}_1^{n+1},\omega_1^{n+1};\dots;\phi_k^{n+1},\hat{c}_k,\omega_k^{n};\dots;\phi_K^{n},\hat{c}_K^{n},\omega_K^{n}\right), \\
	F_{\omega_k}^{n+1}(\omega_k)
	&\equiv& F\left(\phi_1^{n+1},\hat{c}_1^{n+1},\omega_1^{n+1};\dots;\phi_k^{n+1},\hat{c}_k^{n+1},\omega_k;\dots;\phi_K^{n},\hat{c}_K^{n},\omega_K^{n}\right),
\end{eqnarray}
where the superscript $(\cdot)^n$ denotes the current iteration step.
Then, the original joint problem, i.e., the minimization over $\phi_k$, $\hat{c}_k$, and $\omega_k$, is decomposed into a sequence of iterative sub-optimizations
\begin{eqnarray}
	\phi_k^{n+1}&=&\underset{\phi_k\in\Phi}{\operatorname{argmin}}\left\{F_{\phi_k}^{n+1}(\phi_k)\right\}, \\
	\hat{c}_k^{n+1}&=&\underset{\hat{c}_k\in \hat{C}}{\operatorname{argmin}}\left\{F_{\hat{c}_k}^{n+1}(\hat{c}_k)\right\}, \\
	\omega_k^{n+1}&=&\underset{\omega_k\in\mathbb{R}^+}{\operatorname{argmin}}\left\{F_{\omega_k}^{n+1}(\omega_k)\right\},
	\label{eqn:update}
\end{eqnarray}
which can be solved by the calculus of variations.

The corresponding functional for the sub-problem of the spatial distribution $\phi_k(\boldsymbol{x})$ is
\begin{equation}
	J[\phi_k(\boldsymbol{x})]\equiv\int_\Omega\int_{0}^\infty \bigg|\hat{q}(\boldsymbol{x},\omega)-\sum_{i=1}^{k-1}\phi_i^{n+1}(\boldsymbol{x})\hat{c}_i^{n+1}(\omega)-\phi_{k}(\boldsymbol{x})\hat{c}_k^{n}(\omega)-\sum_{i=k+1}^{K}\phi_{i}^{n}(\boldsymbol{x})\hat{c}_i^{n}(\omega)\bigg|^2\mathrm{d}\omega\mathrm{d}\boldsymbol{x}.
\end{equation}
To simplify the formulation above, we define the residual function for the $k$-th mode at iteration step $n$
\begin{equation}
	\hat{r}^n_k(\boldsymbol{x},\omega)=\hat{q}(\boldsymbol{x},\omega)-\sum_{i=1}^{k-1}\phi_i^{n+1}(\boldsymbol{x})\hat{c}_i^{n+1}(\omega)-\sum_{i=k+1}^{K}\phi_{i}^{n}(\boldsymbol{x})\hat{c}_i^{n}(\omega).
\end{equation}
Let the functional derivative vanish for all variations $\phi_k(\boldsymbol{x})+\delta\psi(\boldsymbol{x})\in\Phi,\delta\in\mathbb{R}$, where $\psi$ is an arbitrary function. Then, we get the necessary condition for extrema
\begin{equation}
	\left.\frac{\mathrm{d}}{\mathrm{d}\delta}\left[\int_\Omega\int_{0}^\infty \bigg|\hat{r}^n_k(\boldsymbol{x},\omega)-\phi_{k}(\boldsymbol{x})\hat{c}_k^{n}(\omega)-\delta\psi(\boldsymbol{x})\hat{c}_k^n(\omega)\bigg|^2\mathrm{d}\omega\mathrm{d}\boldsymbol{x}\right]\right|_{\delta=0}=0,
\end{equation}
which further leads to the following equation
\begin{equation}
	0=\int_\Omega\Real\left\{\int_{0}^\infty \overline{\hat{c}_k^n(\omega)}\left[\hat{r}^n_k(\boldsymbol{x},\omega)
	-\phi_{k}(\boldsymbol{x})\hat{c}_k^{n}(\omega)\right] \mathrm{d}\omega\right\}\psi(\boldsymbol{x})\mathrm{d}\boldsymbol{x}.
\end{equation}
Since $\psi(\boldsymbol{x})$ is an arbitrary function, the fundamental lemma of calculus of variations implies that the rest part in the integrand should be identically zero, i.e.,
\begin{equation}
	\Real\left\{\int_{0}^\infty \overline{\hat{c}_k^n(\omega)}\hat{r}^n_k(\boldsymbol{x},\omega)\mathrm{d}\omega\right\}=\int_0^\infty\left|\hat{c}_k^n(\omega)\right|^2\mathrm{d}\omega\ \phi_k(\boldsymbol{x}).
\end{equation}
Let the $L^2$-norm of $\phi_k(\boldsymbol{x})$ be unit, we finally get the update with respect to $\phi_k(\boldsymbol{x})$
\begin{equation}
	\phi_k^{n+1}(\boldsymbol{x})=\frac{\Real\left\{\int_{0}^\infty \overline{\hat{c}_k^n(\omega)}\hat{r}^n_k(\boldsymbol{x},\omega)\mathrm{d}\omega\right\}}{\left\|\Real\left\{\int_{0}^\infty \overline{\hat{c}_k^n(\omega)}\hat{r}^n_k(\boldsymbol{x},\omega)\mathrm{d}\omega\right\}\right\|_{\boldsymbol{x}}}.
	\label{eqn:updatephireal}
\end{equation}

Similarly, the updates with respect to $\hat{c}_k$ and $\omega_k$ become
\begin{eqnarray}
	\hat{c}_k^{n+1}(\omega)&=&\frac{\int_\Omega \phi_k^{n+1}(\boldsymbol{x})\hat{r}^n_k(\boldsymbol{x},\omega)\mathrm{d}\boldsymbol{x}}{1+2\alpha(\omega-\omega_k^n)^2},\quad \omega\geq0\notag\\
	\label{eqn:updatec}\\
	\omega_k^{n+1}&=&\frac{\int_0^\infty\omega\left|\hat{c}_k^{n+1}(\omega)\right|^2\ \mathrm{d}\omega}{\int_0^\infty\left|\hat{c}_k^{n+1}(\omega)\right|^2\ \mathrm{d}\omega},
	\label{eqn:updateomega}
\end{eqnarray}
respectively.

After deriving the sub-problems we actually solve, the complete algorithm in matrix form for computing the RVMD modes is outlined in the following. The flow data, observed from numerical simulations or experimental measurements, is collected as the data matrix
\begin{equation}
	\bm{Q}\in\mathbb{R}^{S\times T},
\end{equation}
where $S$ and $T$ are the numbers of sampling points in space and time, respectively. Note here, since the flow data is generally nonperiodic in time, a mirror extension is conducted in advance to produce the data matrix. The RVMD modes are represented as column vectors
\begin{equation}
	\boldsymbol{\phi}_k\in\mathbb{R}^{S},\quad \boldsymbol{c}_k\in\mathbb{R}^{T}.
\end{equation}

Then the DFT is computed, and we only consider the data matrix and the time coefficient vectors on the non-negative part of the frequency domain, which is indicated using the superscript $(\cdot)^+$ as
\begin{equation}
	\bm{\hat{Q}}^+,\ \boldsymbol{\hat{c}}_k^+.
\end{equation}
The residual matrix for the $k$-th mode is
\begin{eqnarray}
	\bm{\hat{R}}_k^+=\bm{\hat{Q}}^+-\sum_{i\neq k}\boldsymbol{\phi}_i\boldsymbol{\hat{c}}_i^{+,\mathrm{T}},
	\label{eqn:updateresidual_matrix}
\end{eqnarray}
where the superscript $(\cdot)^\mathrm{T}$ denotes the transpose. We then use $(\cdot)^*$ to denote the conjugate and $(\cdot)^\mathrm{H}$ to denote the conjugate transpose of matrices and vectors. Another two diagonal matrices are further defined: the non-negative frequency matrix
\begin{equation}
	\bm{W}^+=\mathrm{diag}\left(\boldsymbol{\omega}^+\right)\quad\text{with}\quad\boldsymbol{\omega}^+=(\omega_i^+)=\left(0,1,2,\dots,T/2\right)\times f_s/T,
\end{equation}
where $f_s$ is the sampling frequency, i.e., the inverse of the sampling interval $1/\Delta t$,
and the filtering matrix
\begin{equation}
	\bm{\hat{G}}_k^+=\mathrm{diag}\left(
	\boldsymbol{\hat{g}}^+
	\right)\quad\text{with}\quad
	\boldsymbol{\hat{g}}^+=(\hat{g}^+_i),\,
	\hat{g}^+_i=\frac{1}{1+2\alpha(\omega^+_i-\omega_k)^2}.
\end{equation}

The superscripts representing the iteration steps, $(\cdot)^n$ and $(\cdot)^{n+1}$, are omitted for simplicity, but each is implicitly understood as the most recent available update.
A C++ implementation of RVMD is available at \href{https://github.com/ZimoLiao/rvmd-cpp}{https://github.com/ZimoLiao/rvmd-cpp}.

\par\vspace{10pt}
\noindent{\bf Algorithm} (Reduced-order variational mode decomposition, RVMD)
\begin{enumerate}[topsep=5pt,leftmargin=20pt,listparindent=0pt,itemindent=0pt,labelsep=5pt]
	\item[(1)] Initialize $\left.\left\{\boldsymbol{\phi}_k,\boldsymbol{\hat{c}}_k^+,\omega_k\right\}\right|_{k=1}^K$ and set iteration step $n\leftarrow0$.
	\item[(2)] Loop:
		\begin{enumerate}[leftmargin=20pt,labelsep=5pt]
			\item[] Set $n\leftarrow n+1$, for $k=1$ to $K$ do:
			\item update the residual matrix for the $k$-th mode using (\ref{eqn:updateresidual_matrix});
			\item update the spatial distribution using
			      \begin{equation}
				      \boldsymbol{\phi}_k^{n+1}=\frac{\Real\left\{\bm{\hat{R}}^+_k\boldsymbol{\hat{c}}_k^{+,*}\right\}}{\left\|\bm{\hat{R}}^+_k\boldsymbol{\hat{c}}_k^{+,*}\right\|};
			      \end{equation}
			\item update the Fourier transform of the time-evolution coefficient using
			      \begin{equation}
				      \boldsymbol{\hat{c}}_k^{+,n+1}=\bm{\hat{G}}^+_k\bm{\hat{R}}^{+,\mathrm{T}}_k\boldsymbol{\phi}_k;
				      \label{eqn:updatec_matrix}
			      \end{equation}
			\item update the central frequency using
			      \begin{equation}
				      \omega_k^{n+1}=\frac{\boldsymbol{\hat{c}}_k^{+,\mathrm{H}}\bm{W}^+\boldsymbol{\hat{c}}_k^{+}}{\boldsymbol{\hat{c}}_k^{+,\mathrm{H}}\boldsymbol{\hat{c}}_k^{+}};
			      \end{equation}
		\end{enumerate}
		Until convergence:
		\begin{equation}
			\sum_{k=1}^K\frac{\|\boldsymbol{\hat{u}}_k^{n+1}-\boldsymbol{\hat{u}}_k^n\|_\mathrm{F}^2}{\|\boldsymbol{\hat{u}}_k^n\|_\mathrm{F}^2}<\epsilon,\quad \boldsymbol{\hat{u}}_k^n\equiv\boldsymbol{\phi}_k^n\boldsymbol{\hat{c}}_k^{+,\mathrm{T},n},
		\end{equation}
		i.e., the iteration difference (the left-hand side) is less than a given tolerance $\epsilon$.
	\item[(3)] Reconstruct the time-evolution coefficient $\boldsymbol{c}_k$ from $\boldsymbol{\hat{c}}_k^+$ using the Hermitian symmetry and inverse Fourier transform.
	\item[(4)] Obtain the RVMD modes $\left.\left\{\boldsymbol{\phi}_k,\boldsymbol{c}_k,\omega_k\right\}\right|_{k=1}^K$.
\end{enumerate}

\subsection{Derived eigenvalue problem from RVMD}
\label{sec:problem_eigen}
As known, the constrained optimization problem of POD leads to an equivalent eigenvalue problem, which can be easily solved. Next, we will show that a similar eigenvalue problem can be derived from the expressions presented above, revealing how the filtering procedure is accomplished in RVMD. Suppose that the convergence has been achieved, i.e.,
\begin{equation}
	\phi_k^{n+1}=\phi_k^n,\hat{c}^{n+1}_k=\hat{c}^{n}_k,\ \omega_k^{n+1}=\omega_k^n.
\end{equation}
We have the filtering function
\begin{equation}
	\hat{g}_k(\omega)\equiv1/[1+2\alpha(|\omega|-\omega_k)^2],
	\label{eqn:filterextension}
\end{equation}
corresponding to a matrix form $\bm{\hat{G}}_k$.
In (\ref{eqn:filterextension}), an even extension is carried out to define the filtering function in the whole frequency domain. So multiplying it with the Fourier transform of a real-valued function does not change the Hermitian symmetry of the function.

An alternative to (\ref{eqn:updatephireal}) can be formulated as
\begin{equation}
	\phi_k^{n+1}(\boldsymbol{x})=\frac{\int_{-\infty}^\infty \overline{\hat{c}^n_k(\omega)}\hat{r}^n_k(\boldsymbol{x},\omega)\mathrm{d}\omega}
	{\left\|\int_{-\infty}^\infty \overline{\hat{c}^n_k(\omega)}\hat{r}^n_k(\boldsymbol{x},\omega)\mathrm{d}\omega\right\|_{\boldsymbol{x}}},
	\label{eqn:updatephi}
\end{equation}
using the Hermitian symmetry in the spectral domain of the time coefficients and the observed space-time data, i.e., $\overline{\hat{c}(\omega)}=\hat{c}(-\omega)$ and $\overline{\hat{q}(\boldsymbol{x},\omega)}=\hat{q}(\boldsymbol{x},-\omega)$. Then substitute (\ref{eqn:updatec}) into (\ref{eqn:updatephi}), we obtain the following identity
\begin{eqnarray}
	\lambda_k\phi_k(\boldsymbol{x})&=&\int_{-\infty}^\infty \hat{g}_k(\omega)\int_\Omega \phi_k(\boldsymbol{x}')\overline{\hat{r}_k(\boldsymbol{x}',\omega)}\mathrm{d}\boldsymbol{x}'\ \hat{r}_k(\boldsymbol{x},\omega)\mathrm{d}\omega \notag\\
	&=&\int_\Omega \left[\int_{-\infty}^\infty \hat{g}_k(\omega)\hat{r}_k(\boldsymbol{x},\omega)\overline{\hat{r}_k(\boldsymbol{x}',\omega)}\mathrm{d}\omega\right]\phi_k(\boldsymbol{x}')\mathrm{d}\boldsymbol{x}',
\end{eqnarray}
where $\lambda_k$ is equal to the norm of the right hand side. In matrix form, we have
\begin{equation}
	\bm{\hat{R}}_k\bm{\hat{G}}_k\bm{\hat{R}}_k^\mathrm{H}\boldsymbol{\phi}_k=\lambda_k\boldsymbol{\phi}_k\quad\text{with}\quad
	\bm{\hat{R}}_k=\bm{\hat{Q}}-\sum_{i\neq k}\boldsymbol{\phi}_i\boldsymbol{\hat{c}}_i^{\mathrm{T}}.
	\label{eqn:eigen_phi}
\end{equation}
Notice that Parseval's theorem implies that $\bm{\hat{R}}_k\bm{\hat{R}}_k^\mathrm{H}=2\pi\bm{R}_k\bm{R}_k^\mathrm{T}$, i.e., the two-point correlation matrix. Hence, RVMD seeks spatial modes that diagonalize the filtered two-point correlation matrix $\bm{\hat{R}}_k\bm{\hat{G}}_k\bm{\hat{R}}_k^\mathrm{H}$, while the adaptive sparsification in the frequency domain is achieved through an iterative optimization over the central frequencies.

\subsection{Criteria for parameters setting}
\label{sec:problem_criteria}
As aforementioned in \S \ref{sec:problem_eigen}, RVMD leads to an eigenvalue problem (\ref{eqn:eigen_phi}) for each mode. This understanding provides some basic criteria for selecting the number of modes $K$ and the filtering parameter $\alpha$ in RVMD.

\begin{figure}
	\centerline{
		\includegraphics[width=.48\linewidth]{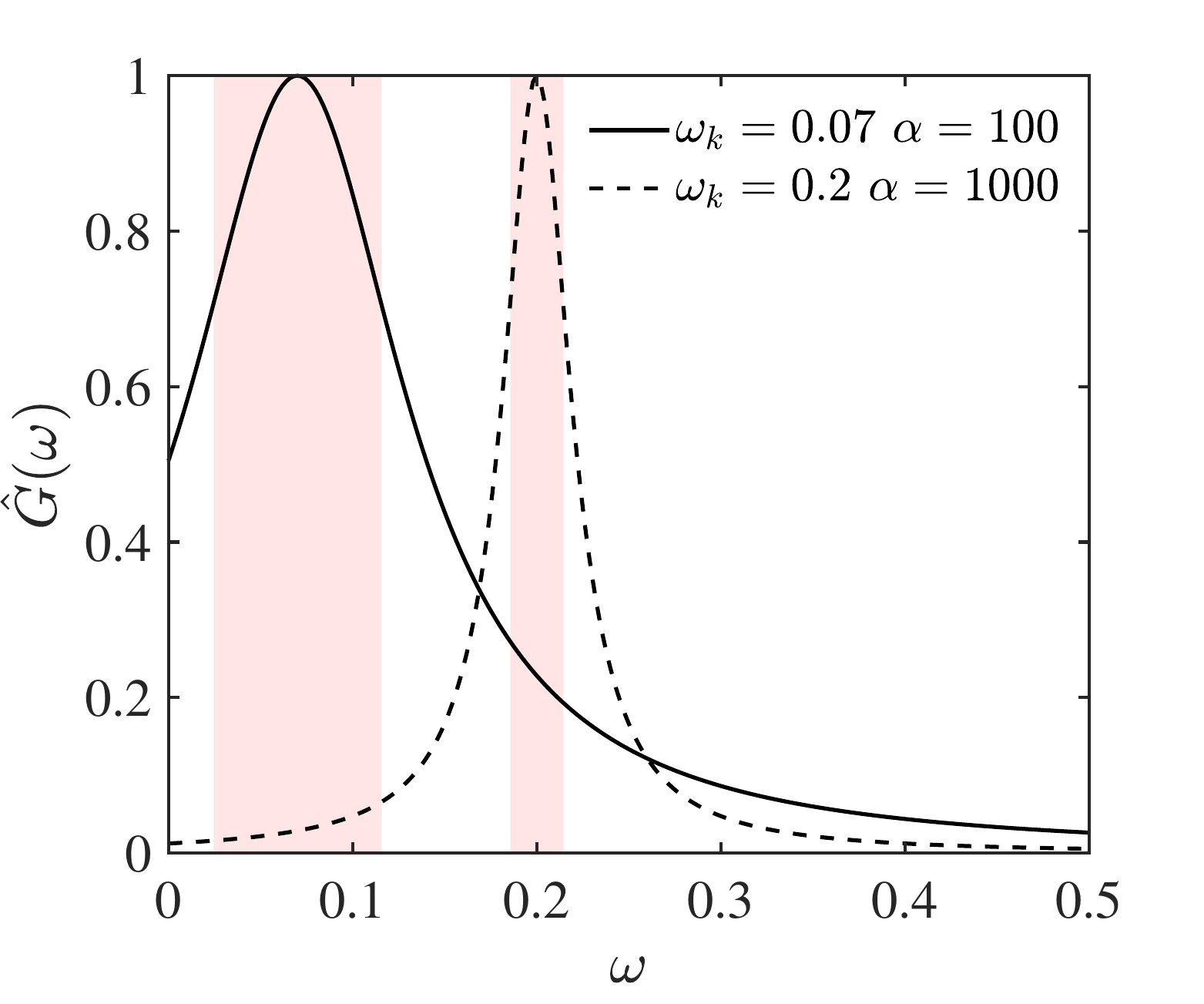}}
	\caption{The filtering functions in RVMD with different filtering parameters $\alpha$ and central frequencies $\omega_k$. The bandwidths of the filters are indicated using a light red background.}
	\label{fig:filter}
\end{figure}

Firstly, consider the most critical input of RVMD -- the filtering parameter $\alpha$. This parameter determines the shape of the filtering function, further affecting the tradeoff between accurate reconstruction and limited bandwidth in RVMD. To estimate an appropriate value of the filtering parameter before performing modal decomposition on specific flow data, we can consider a derived variable, the filter bandwidth $\Delta$, written as
\begin{equation}
	\Delta\equiv\sqrt{(2\sqrt{2}-2)/\alpha}.
\end{equation}
This is the frequency interval between the two half-power points (cutoff frequencies) of the Wiener filter, following the standard definition of the bandwidth of a bandpass filter.
Figure \ref{fig:filter} depicts the filtering functions with different $\alpha$ and $\omega_k$, providing an illustration of the introduced parameters.
Since the space-time data we handle are discrete, the first rule is that the filter bandwidth $\Delta$ should be greater than the frequency resolution of the sampled snapshots. Otherwise, the corresponding $\alpha$ is so large that RVMD is converted into DFT, as we will present in the next section.
Another criterion limiting the upper bound on the filter bandwidth is that $\Delta$ should be small enough to distinguish different dynamics in the flow. If $\Delta$ is too large, e.g., spans the whole frequency domain, the filtered time-evolution coefficient may contain multiple characteristic frequencies like that in POD. Thus, the expected scale separation fails. After the bandwidth's lower and upper bounds are determined, the filtering parameter $\alpha$ can be estimated easily. Secondly, the setting of $K$ is more arbitrary. The number of modes does not alter the adaptivity of RVMD or the dynamical properties of computed modes. So this parameter is merely the expected number of distinctive dynamical behaviours contained in the specific flow.

Finally, since the BCD algorithm can not find the global minimum for the constructed non-convex optimization problem, different central frequencies initialization strategies may lead to different local minimums, i.e., the RVMD modes. When we know very little about the frequency characteristics of the considered flow, initializing the central frequencies to be uniformly or randomly distributed over the entire frequency domain is recommended. In this case, no prior knowledge is required. Furthermore, when a qualitative or quantitative understanding of the frequency characteristics has been established, e.g., through the PSD of the probed temporal sequences or the prediction based on theoretical analysis, a user-defined central frequencies initialization may lead to a better performance -- faster convergence, and a smaller number of modes that is enough to capture the dominant dynamics -- of the proposed method. A detailed parametric study can be found in \S~\ref{sec:rectjet}, practically demonstrating the above discussion on the parameter dependence of RVMD.

\section{Relation to existing methods}
\label{sec:relation}
\begin{figure}
	\centerline{
		\includegraphics[width=.8\linewidth]{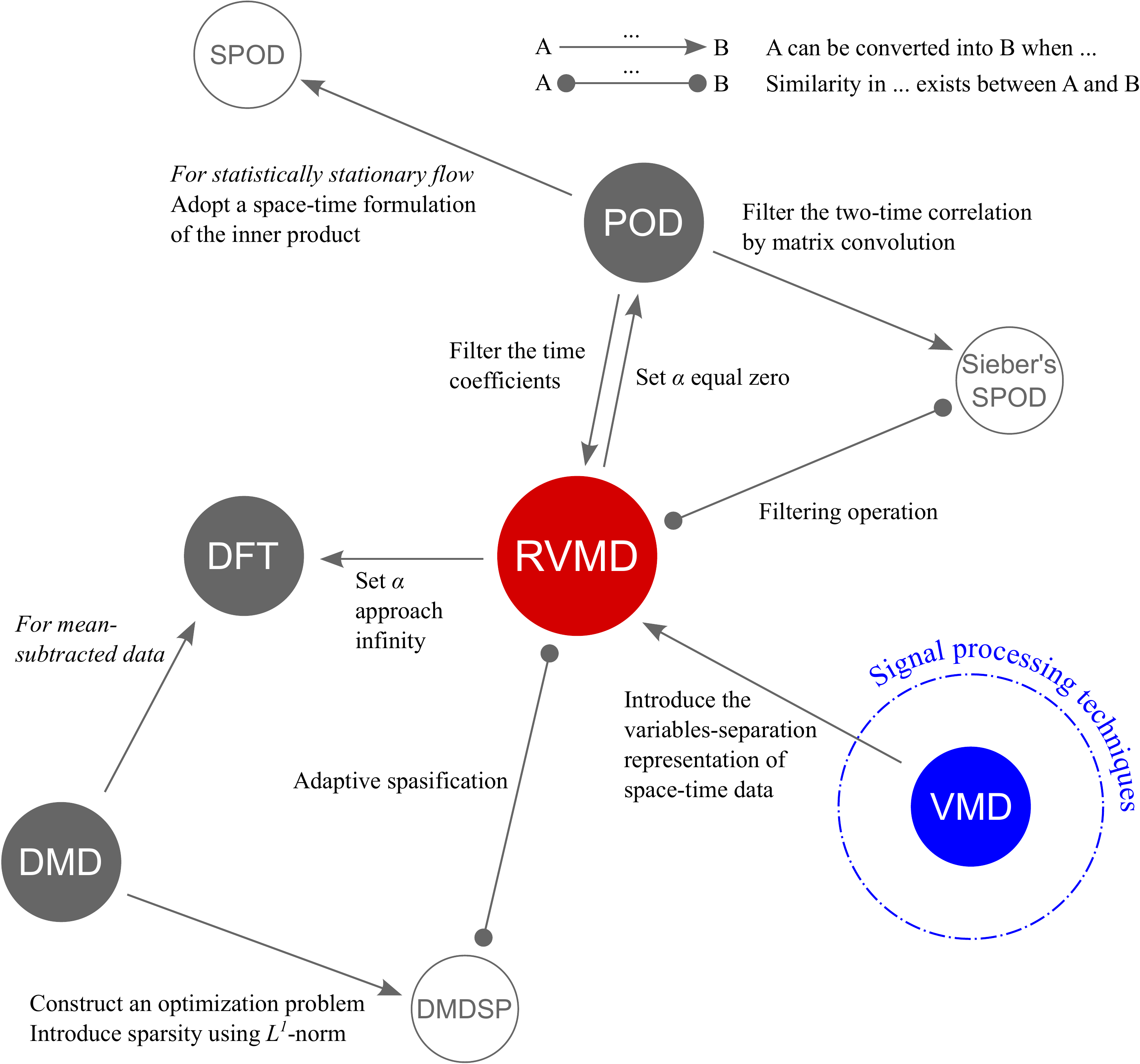}}
	\caption{The general relations between RVMD and other methods.}
	\label{fig:map}
\end{figure}

This section expounds on the relations between RVMD and some classical modal decomposition methods. We show that RVMD can be converted into POD or DFT at extreme parameter selections. Similarities in construction ideas between RVMD and other widely-used methods are also discussed. Furthermore, to illustrate the advantages of RVMD in performing time-frequency analysis, we briefly introduce the nonstationary signal processing techniques and present a signal-processing analogous categorization of modal decomposition methods. The general relations between RVMD and some other methods are summarized in figure \ref{fig:map}.

\subsection{Proper orthogonal decomposition}
As the filtering parameter $\alpha$ is set as zero, $\bm{\hat{G}}_k$ reduces to the identity matrix, then (\ref{eqn:eigen_phi}) turns into
\begin{equation}
	\bm{\hat{R}}_k\bm{\hat{R}}_k^\mathrm{H}\boldsymbol{\phi}_k=2\pi\bm{{R}}_k\bm{{R}}_k^\mathrm{T}\boldsymbol{\phi}_k=\lambda_k\boldsymbol{\phi}_k.
\end{equation}
If each spatial distribution vector is further restricted to be orthogonal to any other one, we obtain the POD modes directly as a consequence of the reduced version of RVMD. This point can be reached in another way. Specifically, if we drop the bandwidth term and apply an orthogonality constraint, the original problem (\ref{eqn:rvmd}) becomes
\begin{eqnarray}
	\min_{\left\{\phi_k(\boldsymbol{x}),c_k(t)\right\}}
	\Bigg\{
	\bigg\|q(\boldsymbol{x},t)-\sum_{k=1}^K\phi_k(\boldsymbol{x})c_k(t)\bigg\|_\mathrm{F}^2
	\Bigg\},\notag\\
	\text{s.t.\quad}\left\langle\phi_i(\boldsymbol{x}),\phi_k(\boldsymbol{x})\right\rangle_{\boldsymbol{x}}=\delta_{ik}.
\end{eqnarray}
This problem is equivalent to the following minimization over the spatial distribution functions
\begin{eqnarray}
	\min_{\left\{\phi_k(\boldsymbol{x})\right\}}
	\Bigg\{
	\bigg\|q(\boldsymbol{x},t)-\sum_{k=1}^K\frac{\langle\phi_k(\boldsymbol{x}),q(\boldsymbol{x},t)\rangle_{\boldsymbol{x}}}{\|\phi_k(\boldsymbol{x})\|_{\boldsymbol{x}}^2}\phi_k(\boldsymbol{x})\bigg\|_\mathrm{F}^2
	\Bigg\},\notag\\
	\text{s.t.\quad}\left\langle\phi_i(\boldsymbol{x}),\phi_k(\boldsymbol{x})\right\rangle_{\boldsymbol{x}}=\delta_{ik},
\end{eqnarray}
which is exactly what POD solves \citep{Holmes12}.

\subsection{Discrete Fourier transformation}
The other extreme case is of $\alpha\to\infty$. In this condition, the only nonzero component of the time coefficient vector $\boldsymbol{\hat{c}}_k^+$ is exactly the one with the central frequency $\omega_k$, as indicated by (\ref{eqn:updatec_matrix}). In other words, when the filtering parameter $\alpha$ approaches infinity, the time coefficients are pure harmonics, indicating that the RVMD modes reduce to the Fourier modes.

\subsection{Other methods and a signal-processing analogous categorization}
We have shown that the proposed method would be strictly converted into POD and DFT in two extreme cases. Besides, it is worth noting that RVMD also shares some similarities in the physical considerations or the mathematical operations with various newly developed modal decomposition methods. In the following, we will discuss the relations between RVMD and other methods. Then, a signal-processing analogous categorization is presented to illustrate the capability of each technique.

Since we introduce a filtering term in the constructed optimization problem to obtain limited-bandwidth time coefficients, the optimization procedure finally results in an eigenvalue problem of the filtered two-point correlation matrix (\ref{eqn:eigen_phi}). A similar idea is adopted in \citet{Sieber16}. Here, we call the method proposed in \citet{Sieber16} as Sieber's SPOD. Based on the experimental observations on the two-time correlation, Sieber's SPOD computes eigenmodes of a filtered two-time correlation matrix and then obtains time coefficients similar to IMFs. In Sieber's SPOD, the filtering is conducted through a matrix convolution, while RVMD filters the time coefficients directly. Thus, compared to Sieber's SPOD, RVMD presents a mathematically well-defined framework with clearer physical interpretability.

Another essential feature of RVMD is the adaptivity in frequency-domain sparsification. This feature was previously discussed in \citet{Jovanovic14}, in which a sparsity-promoting variant of DMD called DMDSP was proposed. In DMDSP, the sparsity is induced by an additional penalty term -- $L^1$-norm of the vector of DMD amplitudes, while a $L^2$-square norm of the gradient of time coefficient is introduced in RVMD to realize this feature. Similarly, RVMD and DMDSP handle a constructed optimization problem to achieve a tradeoff between accurate reconstruction and low-redundant sparse representation.

\begin{table}
	\centering
	\def~{\hphantom{0}}
	\begin{tabular}{lcc}
		\toprule
		Category                                 & Signal processing & Modal decomposition \\[2ex]
		Time-domain analysis                     &                   & POD                 \\[1.5ex]
		Frequency-domain analysis                & FT                & DMD, SPOD           \\[1.5ex]
		\multirow{2}{*}{Time-frequency analysis} & STFT, WT          & mrDMD               \\
		                                         & EMD (HHT),  VMD   & RVMD (present)      \\
		\bottomrule
	\end{tabular}
	\caption{A signal-processing analogous categorization of modal decomposition methods}
	\label{tab:categorization}
\end{table}

According to the ability to describe temporal evolution, the existing modal decomposition methods can be generally divided into three categories. This signal-processing analogous categorization is listed in table \ref{tab:categorization}. Since both DMD and SPOD modes oscillate at a single frequency, they can be seen as space-time extensions of the Fourier transform (FT). As we have mentioned above, Fourier-based methods such as short-time Fourier transform (STFT) and wavelet transform (WT) provide frameworks to carry out time-frequency analysis. The multiresolution DMD (mrDMD) lies in this hierarchy, which introduces the idea of wavelet-based multiscale analysis. Although the completeness of mathematics makes the Fourier-based methods widely used and deeply studied, they have to encounter a defect in determining how to select the bases properly as well as the resolution limitation due to the uncertainty principle (also referred to as the Gabor limit). To this end, \citet{Huang98} proposed a new methodology called Hilbert-Huang transform (HHT) to analyze nonstationary and nonlinear time series in the Hilbert view. The standard HHT decomposes a signal into IMFs using empirical mode decomposition (EMD) and then applies a Hilbert spectral analysis on IMFs to get the instantaneous frequencies. VMD acts as an alternative mathematically well-defined approach for extracting IMFs and performs better than EMD. Since RVMD is a space-time extension of VMD, it inherits the advantages of VMD/EMD in time-frequency representation. RVMD provides an approximate variables-separation representation for space-time data, in which the time-evolution coefficients are IMFs. Based on the idea of HHT, a Hilbert spectral analysis (HSA) can further be performed on the computed RVMD time coefficients, leading to the space-time-frequency characteristics of the original data.

\section{Applications}
\label{sec:application}
In this section, we apply RVMD to two examples of canonical flow problems: the transient cylinder wake and the rectangular turbulent supersonic screeching jet. The first example is intended to practically introduce the time-frequency analysis framework based on RVMD. The second example is included to demonstrate the applicability of RVMD to complex turbulent flows. The parameter dependence of RVMD is discussed in the latter case to verify the statements we have made in \S~\ref{sec:problem_criteria}.

\subsection{The transient cylinder wake}
\begin{figure}
	\centerline{
		\includegraphics[height=.4\linewidth]{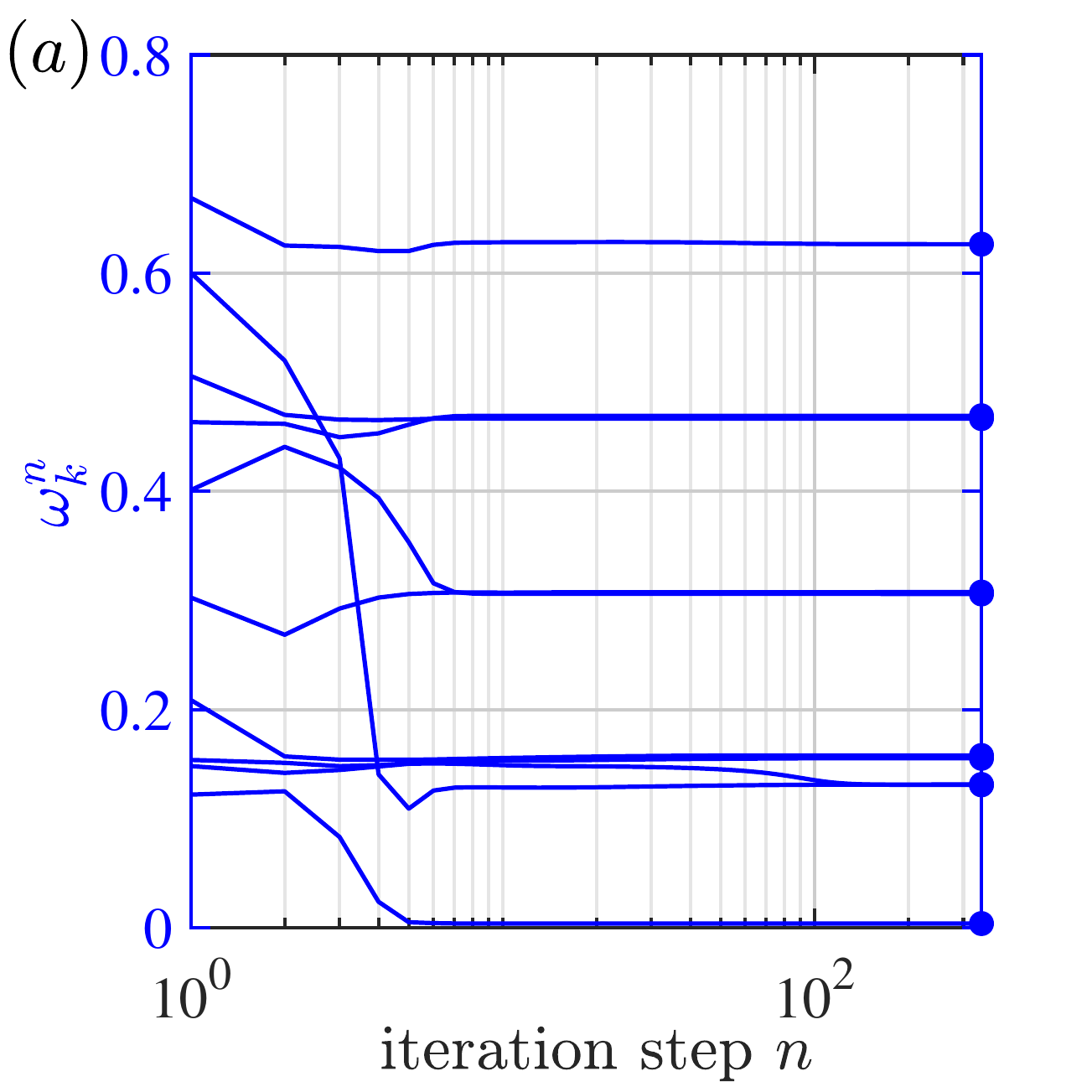}
		\includegraphics[height=.4\linewidth]{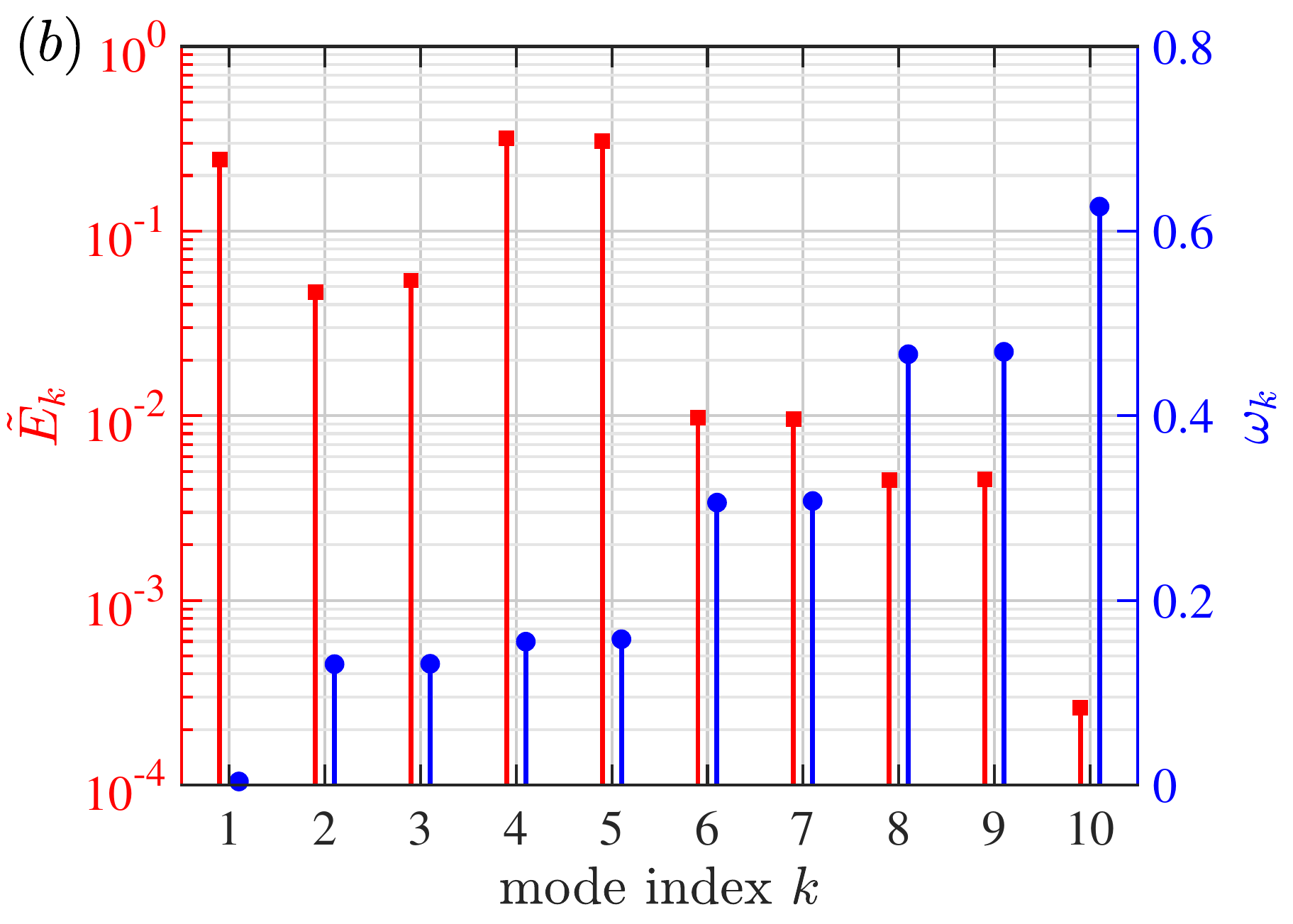}}
	\caption{RVMD results for the transient cylinder wake: ($a$) convergence curve of central frequencies, ($b$) energy ratios ($\tilde{E}_k$) and central frequencies ($\omega_k$). For clarity, small shifts are applied to the abscissas in panel $b$ for the plots of $\tilde{E}_k$ and $\omega_k$. }
	\label{fig:cylinderrvmd}
\end{figure}

We begin with considering the transient cylinder wake, a widely-used problem to validate low-dimensional modelling \citep{Noack03, Brunton16}. A two-dimensional incompressible flow past a cylinder is simulated using a lattice Boltzmann solver. The flow variables are non-dimensionalized by the cylinder diameter $D$, the inflow velocity $U$, and the kinetic viscosity $\nu$. We set the Reynolds number $\Rey=UD/\nu$ equal to $100$ to ensure the onset of periodic vortex shedding \citep{Zebib87}, similar to the numerical setups in \citet{Noack16RDMD}. The origin of coordinates coincides with the center of the cylinder, and the computational domain is set as $(x,y)\in[-10,15]\times[-10,10]$. A uniform inflow condition $u=1,v=0$ is implemented at the inlet and transverse boundaries, and an extrapolation condition at the outlet boundary. The no-slip boundary condition at the cylinder is imposed by immersed boundary method.

In this simulation, the flow is initially set as the unstable steady state and then gradually evolves to a periodic vortex shedding -- known as the von K\'arm\'an vortex street, which can be restated in the language of dynamical systems as a transition from the unstable fixed point to the stable limit-cycle oscillation \citep{Zebib87}. To include the whole transient process and ensure the time resolution high enough, $600$  snapshots of the mean-subtracted velocity fields with sampling interval $\Delta t U/D=0.25$ are collected to perform modal decomposition. In this case, we set the mode number $K=10$ and the filtering parameter $\alpha=1000$, and initialize the central frequencies to be uniformly distributed in the interval $[0,0.6]$. The setting of empirical parameters follows the criteria proposed in \S~\ref{sec:problem_criteria}. The iterative optimization reaches convergence after 343 steps, with the iteration difference below $0.2\%$, the convergence curve of the central frequencies $\omega_k^n$ are depicted in figure \ref{fig:cylinderrvmd}($a$). The RVMD modes obtained are indexed by their central frequencies from low to high.

\begin{figure}
	\centerline{
		\includegraphics[width=.96\linewidth]{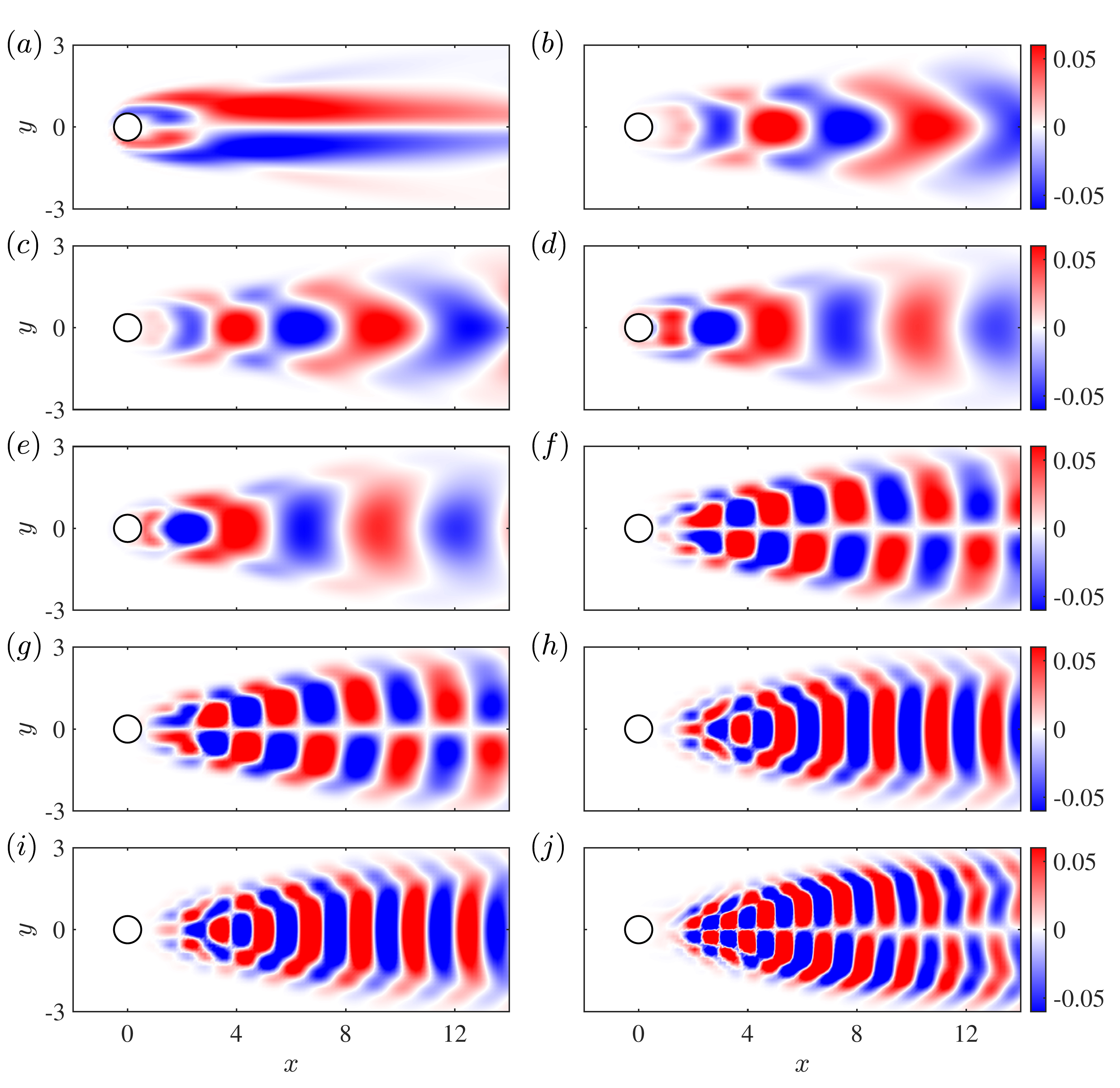}}
	\caption{Spatial distributions of the RVMD modes (shown by vorticity) for the transient cylinder wake. We perform a zoom-in to concentrate on the near-wake region plotted for ($a$-$j$) modes 1-10.}
	\label{fig:cylinderphi}
\end{figure}

\begin{figure}
	\centerline{
		\includegraphics[width=.96\linewidth]{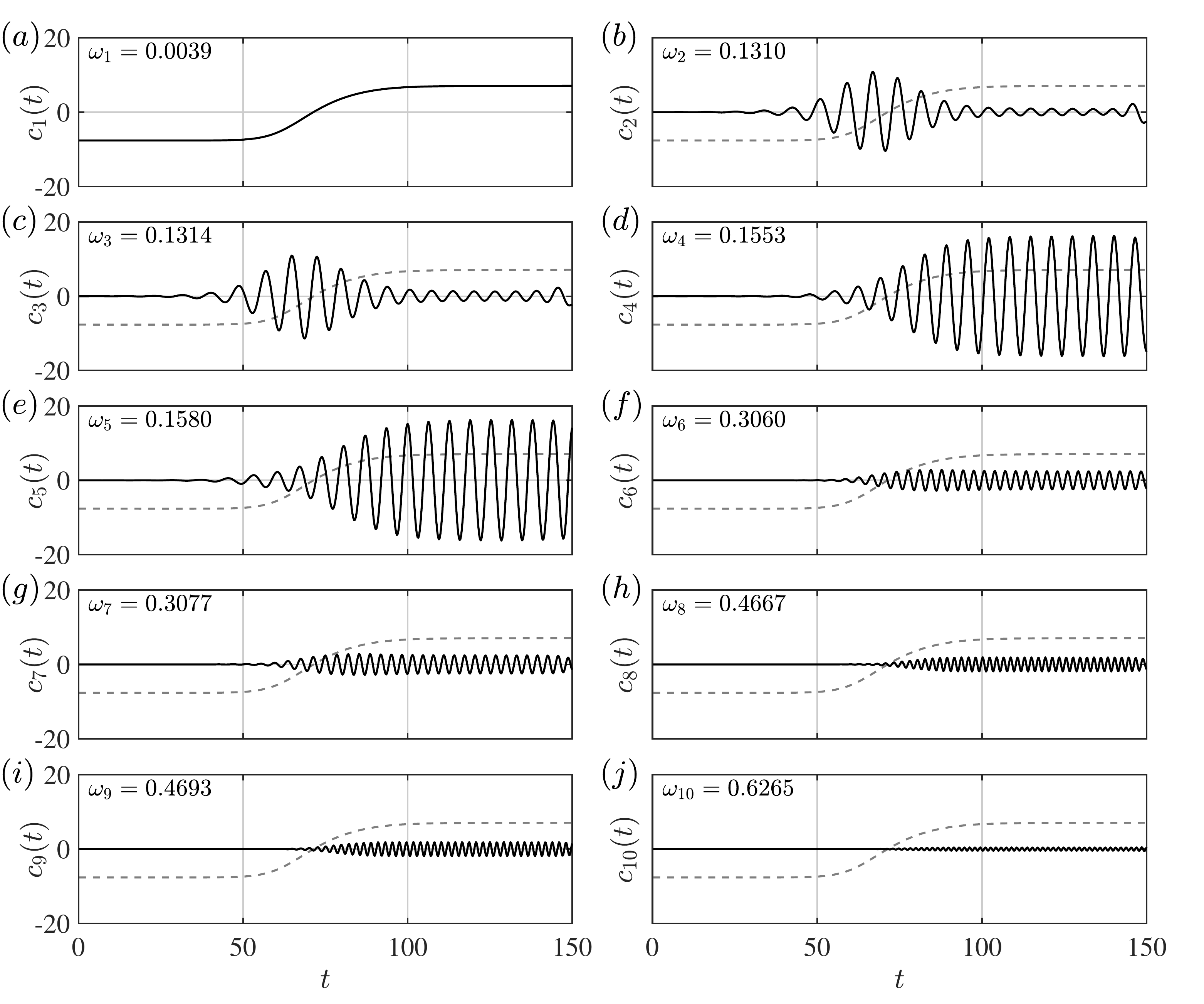}}
	\caption{Time-evolution coefficients of the RVMD modes for the transient cylinder wake. ($a$-$j$) Modes 1-10. The gray dashed curve indicates the temporal evolution of the shift-mode.}
	\label{fig:cylinderc}
\end{figure}

A detailed comparative study on different modal decomposition methods has been carried out by \citet{Noack16RDMD}, which shows that both POD and DMD can not deal with this problem properly. Since DMD can only capture oscillation modes with exponential envelopes, the intermediate transient process is wiped, indicating a wrong evolutionary trend. POD performs relatively better in extracting the temporal behaviour yet may mix multiple frequencies, leading to confusion in the physical meaning of the modes obtained. The spatial distributions and the corresponding time-evolution coefficients of the RVMD modes are depicted in figures \ref{fig:cylinderphi} and \ref{fig:cylinderc}, respectively. The computed modes can be grouped into pairs -- the oscillation frequencies and energies of the two modes paired are almost the same, but their phases are different, which is consistent with the previous modal analysis of cylinder wakes \citep{Bagheri13}.
The post-transient von K\'arm\'an vortex street is precisely captured by the first harmonics (modes 4,\,5), the second harmonics (modes 6,\,7), the third harmonics (modes 8,\,9), and the fourth harmonics (mode 10). Mode 1 corresponds to the so-called shift-mode reported by \citet{Noack03}, defined as the transient from the initially steady solution to the post-transient time-averaged flow. It should be emphasized that due to the limited-bandwidth property of the RVMD modes, there is no spurious oscillation mixed in the temporal evolution of the shift-mode (see figure \ref{fig:cylinderc}($a$)).

\begin{figure}
	\centerline{
		\includegraphics[height=.4\linewidth]{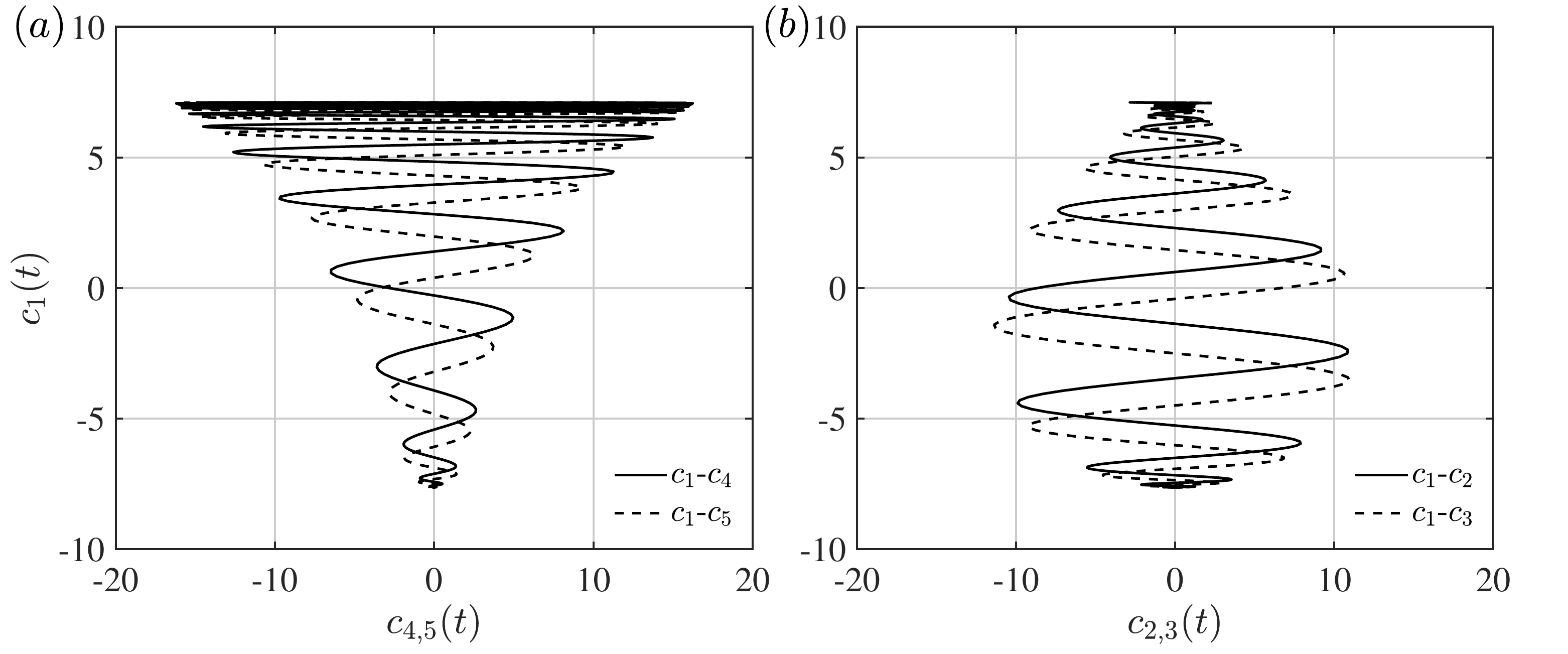}}
	\caption{The transient trajectory in RVMD coordinates. The shift-mode versus ($a$) the periodic vortex shedding modes, ($b$) the transient linear stability modes.}
	\label{fig:cylindertrajectory}
\end{figure}

\begin{figure}
	\centerline{\includegraphics[height=.4\linewidth]{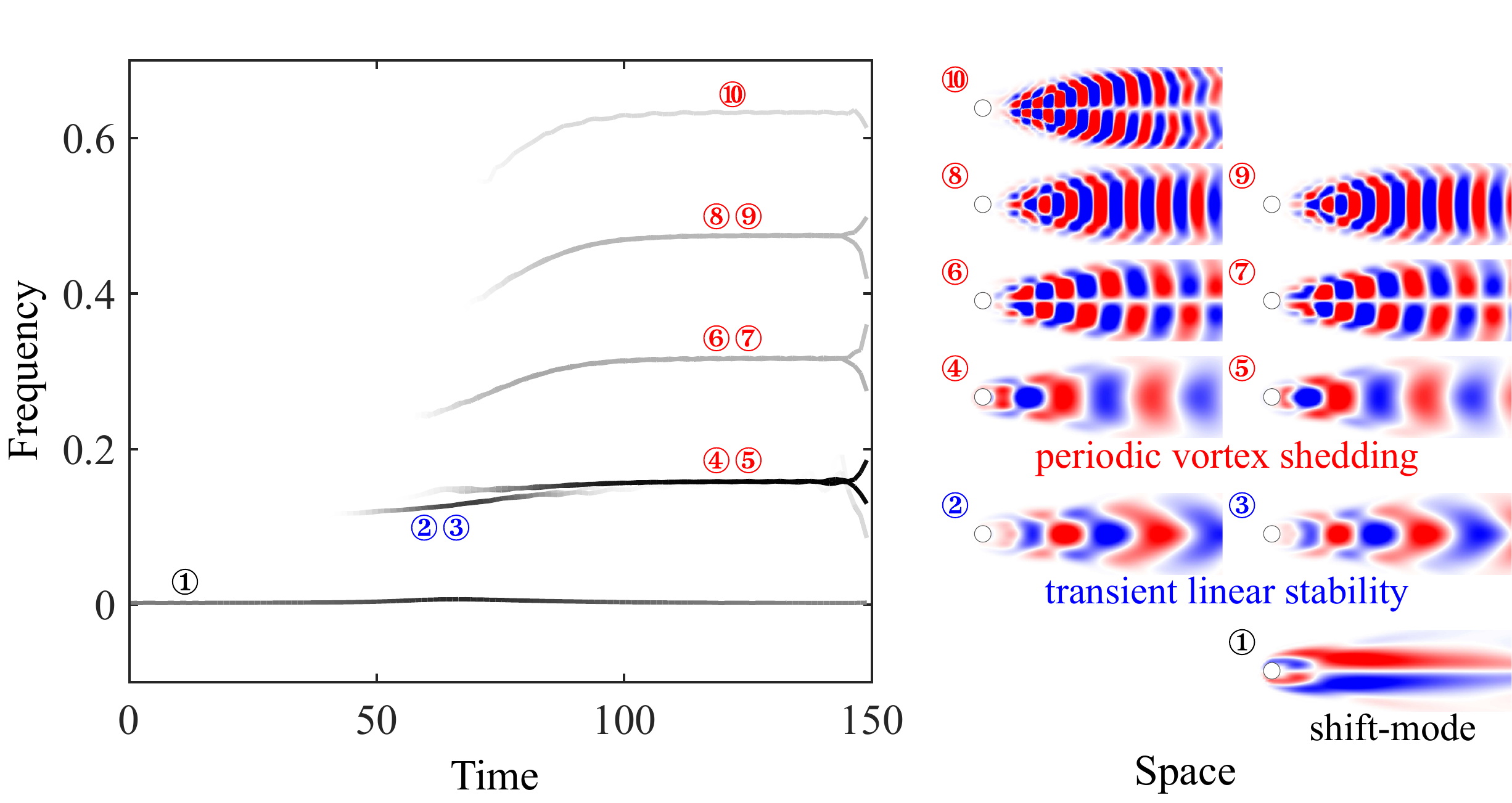}}
	\caption{The Hilbert spectrum of the time-evolution coefficients and the corresponding spatial distributions. In the Hilbert spectrum, the curve of each RVMD mode is grey-scaled by its energy that evolves in the time-frequency plane, with darker for higher energy. At the end of the time axis, an abrupt increase or decrease of frequency arises for each RVMD mode due to the end effects inherent in computing the Hilbert spectrum \citep{Huang98}, other than the physical evolution dynamics.}
	\label{fig:cylinderhsa}
\end{figure}

Another intriguing finding is that RVMD resolves the intermediate vortex shedding, as shown in figure \ref{fig:cylinderphi}($b,\,c$) and figure \ref{fig:cylinderc}($b,\,c$). The mode pair's central frequency $\omega_{2,3}\simeq 0.131$ matches the results of linear stability analysis \citep{Noack03}, in which the predicted frequency is $0.135$. Two significant differences exist between the transient linear stability modes (modes 2,\,3) and the periodic vortex shedding modes (modes 4,\,5). First, the maximum fluctuation of transient linear stability modes is far from the cylinder, while the maximum fluctuation of vortex shedding modes nears the cylinder, as shown in figure \ref{fig:cylinderphi}($b$-$e$). Second, while the transient linear stability modes experience a growing-peaking-decaying process, the vortex shedding modes have a later onset and finally reach a stable-oscillation state. As illustrated in figure \ref{fig:cylindertrajectory}, the trajectories of the transient cylinder wake in RVMD coordinates, i.e., curves $c_1$-$c_{4,5}$ and $c_1$-$c_{2,3}$ display totally different evolution characteristics. This observation indicates that the distinguishment of the two behaviours may be critical to describing the whole flow process, further affecting the predictive power of the constructed low-dimensional model \citep{Noack03}.
Compared to existing methods \citep{Siegel08,Noack16RDMD}, RVMD can provide an empirical expansion for the transient cylinder wake with the best physical interpretability. Each distinctive behaviour -- namely shift-mode, transient linear stability modes, and periodic vortex shedding modes -- is captured and isolated, avoiding frequency mixing. These results further confirm the advantages of spatially non-orthogonal modes in capturing dynamical characteristics.

After extracting the IMF-like time coefficients by RVMD, we can further perform a Hilbert spectral analysis to get the space-time-frequency characteristics of the transient cylinder wake. The analytic representations of the real-valued time coefficients $c_k(t)$ are considered as
\begin{equation}
	c_{A,k}(t)=c_k(t)+\mathrm{i}\mathcal{H}c_k(t)=A_k(t)e^{\mathrm{i}\varphi_k(t)},
\end{equation}
where the operator $\mathcal{H}$ denotes the Hilbert transform, $A_k(t)$ is the envelope, and $\varphi_k(t)$ is the phase (see appendix \ref{app:analytic}). The instantaneous frequency is defined as $\mathrm{d}\varphi_k(t)/\mathrm{d}t$. The Hilbert spectrum is depicted in figure \ref{fig:cylinderhsa}. By combining RVMD and HSA, we get a picture that indicates the temporal dynamics in a physically intuitive way. A one-to-one correspondence exists between the spatial distribution and the evolution curve in the time-frequency plane of each RVMD mode, providing a sparse and powerful description of the transient dynamics.

\subsection{The rectangular turbulent supersonic screeching jet}
\label{sec:rectjet}
\subsubsection{Flow configuration}
Here, we take the rectangular supersonic screeching jet as a typical example to demonstrate the applicability of RVMD in practical turbulent flows. Since \citet{Powell53} first detected the discrete-frequency high-energy screech in underexpanded supersonic jets, many efforts have been devoted to describing, explaining, and predicting this phenomenon. Understanding the key characteristics and the underlying mechanisms in screeching jets is crucial for aeroacoustic theory and aerospace engineering. However, the complex flow processes -- shear-induced turbulence, shock cells, and their nonlinear interactions -- make it hard to reveal the origin of screech and present reliable predictions. In recent years, data-driven modal analysis has provided a new perspective to quantify and interpret this phenomenon \citep{Jovanovic14,Li21,Edgington22}.

\begin{figure}
	\centerline{\includegraphics[height=.3\linewidth]{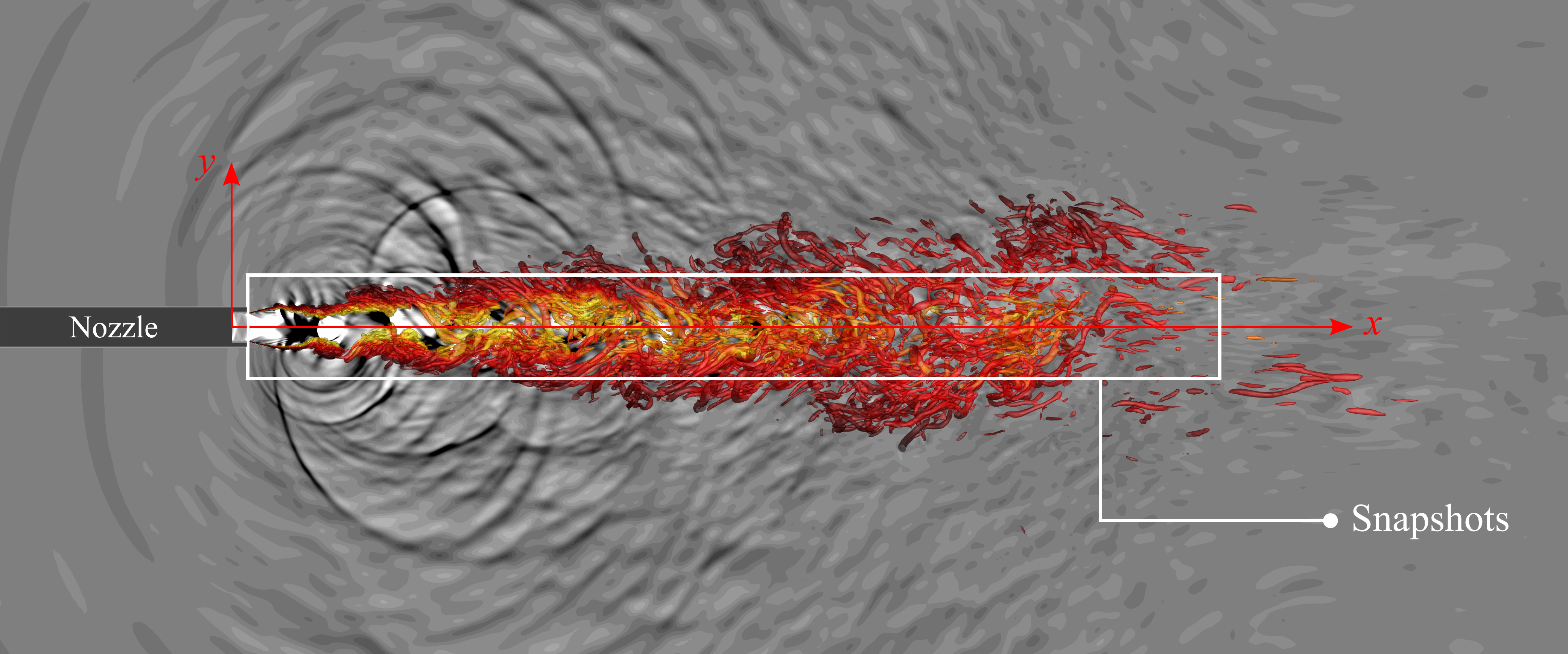}}
	\caption{Flow configuration of the rectangular turbulent supersonic screeching jet. The vortex structures are shown by the $Q$-criterion (colored by the streamwise velocity), and the divergence of velocity is contoured as the background.}
	\label{fig:rectjetconfig}
\end{figure}

We use data obtained via a high fidelity implicit large eddy simulation (ILES) on a Cartesian grid \citep{Ye20}. This ILES was performed using an in-house compressible flow solver HiResX. All the variables are non-dimensionalized using the nozzle height $h$, the fully expanded jet velocity $U_j$, the far-field pressure $p_\infty$ and density $\rho_\infty$, and the reference dynamic viscosity $\mu_\infty$. The jet operates at the nozzle pressure ratio $NPR=2.09$ and a fully-expanded Mach number $M_j\equiv U_j/a_j=1.55$. The far-field sound speed and the jet sound speed are $a_\infty\simeq339\ \mathrm{m/s}$ and $a_j\simeq278\ \mathrm{m/s}$, respectively. Since the spanwise boundaries of this simulation are periodic, the three-dimensional effect is relatively weak. The calculated shock-cell spacing ($L_s\simeq2.51h$) and the fundamental screech tone (Strouhal number $St_s=\omega_s h/U_j=0.114$) agree well with previous experimental measurements \citep{Raman94,Panda97} and LES data \citep{Berland07}. Details on numerical setups can be found in \citet{Ye20}. 1600 snapshots of the fluctuating pressure are sampled in the streamwise-transverse plane with a dimensionless time interval $\Delta t U_j/h=0.1$ to perform modal analysis. A rectangle domain $(x,y)\in[0.6h,35.2h]\times[-1.86h,1.86h]$ is chosen to contain the shock cells and the surrounding shear layers, as shown in figure \ref{fig:rectjetconfig}.

\subsubsection{Test on the parameters setting}
A detailed study on the parameters setting of RVMD is performed first to further verify our statements in \S~\ref{sec:problem_criteria}. We will quantitatively assess the dependence of the RVMD results on the two inputs, i.e., the number of modes $K$ and the filtering parameter $\alpha$, and how the initialization strategies of the central frequencies (corresponding to Strouhal number $St_k\equiv \omega_k h/U_j$ hereafter) affect the RVMD results.

By theory, each RVMD mode can be seen as an eigenvector of the filtered two-point correlation matrix, and the filtering function is determined by $\alpha$ as
\begin{equation}
	\frac{1}{1+2\alpha(\omega-\omega_k)^2}.
\end{equation}
This filter provides a prior spectrum of the time coefficient $c_k(t)$ with bandwidth $\Delta=[(2\sqrt{2}-2)/\alpha]^{1/2}$ around $\omega_k$. To illustrate the dependence on $\alpha$ (i.e., $\Delta$), ten trials are presented at the same number of modes ($K=24$) and the same initialization of the central frequencies (uniformly distributed in the whole frequency domain). The values of the filtering parameters and the corresponding bandwidths are listed in table \ref{tab:param_alpha}.

\begin{figure}
	\center{
		\includegraphics[width=.6\linewidth]{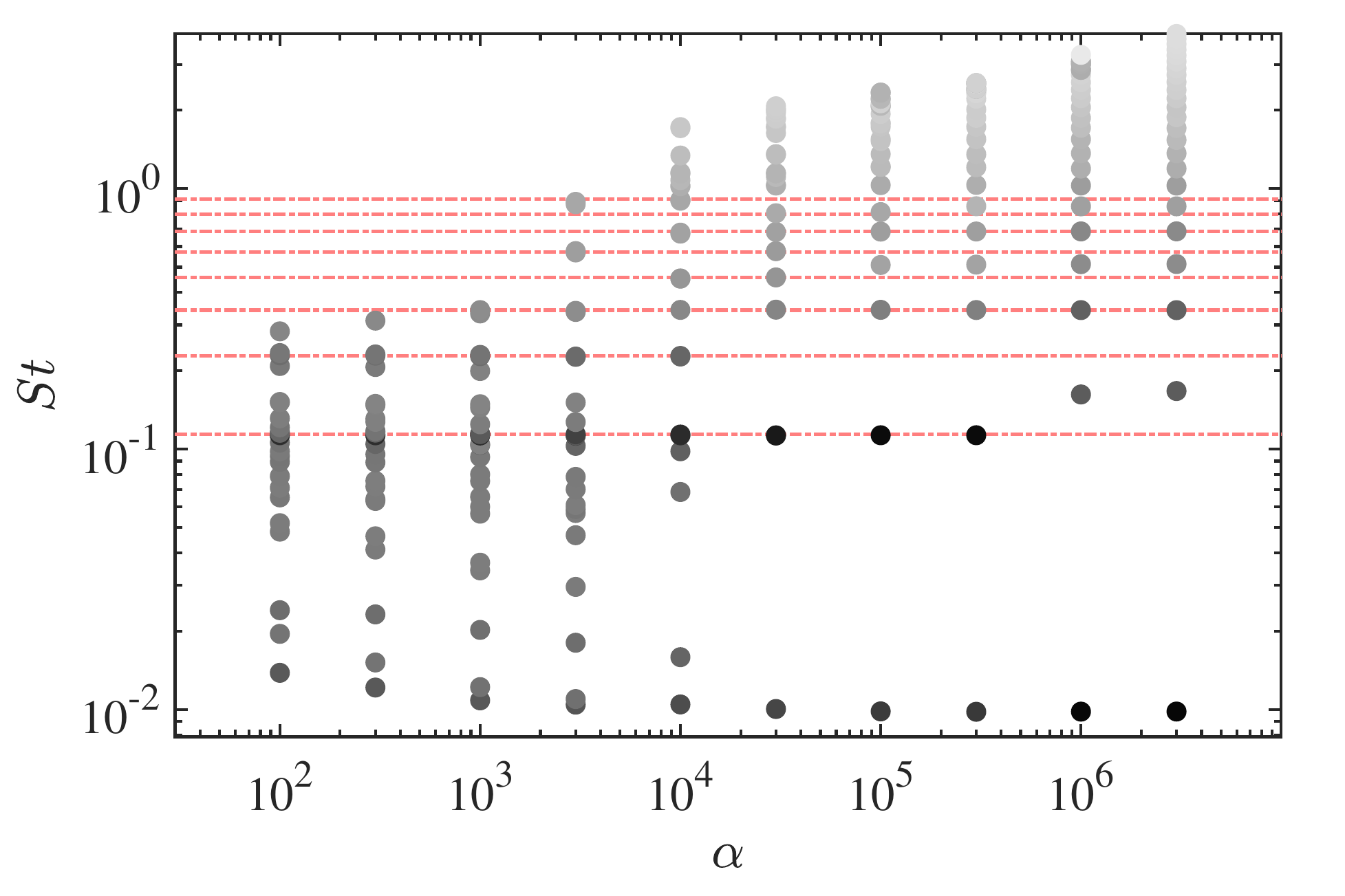}
	}
	\caption{RVMD modes for the rectangular screeching jet at different filtering parameters $\alpha$. The marked point of each RVMD mode is grey-scaled by its energy ratio, i.e., darker for higher energy. The red dash-dot lines indicate the fundamental screech tone $St_s=0.114$ and its harmonics.}
	\label{fig:param_alpha}
\end{figure}

As indicated in figure \ref{fig:param_alpha}, a proper choice of $\alpha$ is crucial for the expected properties of RVMD: the ability to capture finite-bandwidth dynamics and the adaptivity in extracting the characteristic frequencies. For large $\alpha$ (i.e., small $\Delta$), the adaptivity is dropped, and the iterative algorithm finally obtains the Fourier modes at the given initial central frequencies. For small $\alpha$ (i.e., large $\Delta$), RVMD reduces to POD as expected. In this case, the local energetic dominant behaviours at relatively high frequencies are neglected, and a large amount of the RVMD modes represent low-frequency broadband behaviours that are difficult to interpret in flow physics. This adaptive modal decomposition realizes its best performance when we set $\alpha$ around $10000$. RVMD precisely captures the high-energy harmonic modes of jet screeching. Meanwhile, some low-frequency dynamics are extracted by RVMD. We will show later that these modes reveal an oscillatory stretching of the shock cell and may affect the amplitudes of the screeching modes. Although the filtering parameter is indeed flow-problem dependent, it has been shown here that a proper choice can be easily made post hoc, as in VMD \citep{Dragomiretskiy14}.

\begin{table}
	\centering
	\def~{\hphantom{0}}
	\resizebox{1.\linewidth}{!}{
		\begin{tabular}{crrrrrrrrrr}
			\toprule
			$\alpha$          & $100$    & $300$   & $1000$  & $3000$  & $10000$ & $30000$ & $100000$ & $300000$ & $1000000$ & $3000000$ \\[1ex]
			$\Delta$ (points) & $145.63$ & $84.08$ & $46.05$ & $26.59$ & $14.56$ & $8.41$  & $4.61$   & $2.66$   & $1.46$    & $0.84$    \\
			\bottomrule
		\end{tabular}
	}
	\caption{The filtering parameters and the corresponding filter bandwidths (in sampling points).}
	\label{tab:param_alpha}
\end{table}

\begin{figure}
	\centerline{
		\includegraphics[width=.96\linewidth]{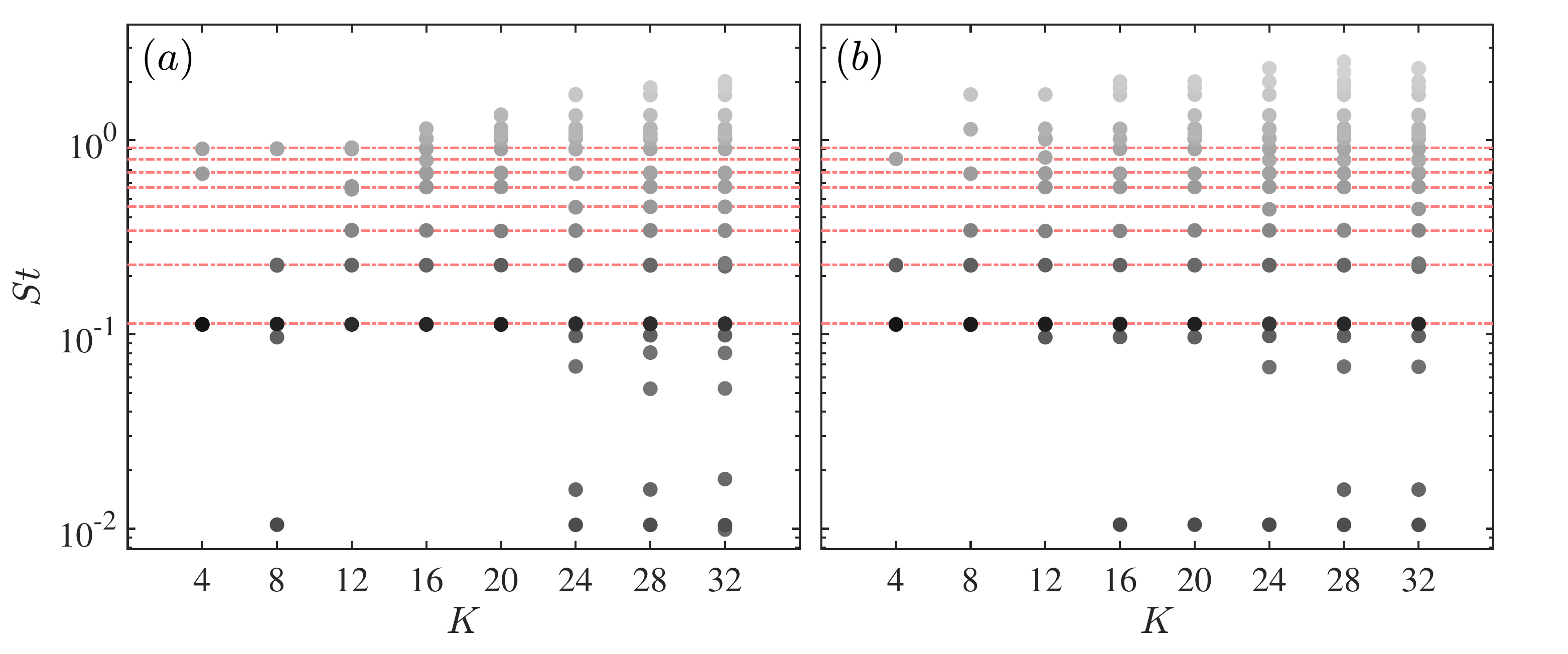}}
	\caption{RVMD modes for the rectangular screeching jet using the same filtering parameters $\alpha=10000$ but different numbers of modes $K$. The central frequencies are initialized to be ($a$) uniformly, ($b$) quadratically distributed in the frequency domain.}
	\label{fig:param_K_init}
\end{figure}

Figure \ref{fig:param_K_init}($a$) displays the results of RVMD with the same filtering parameters ($\alpha=10000$), the same initialization of the central frequencies (uniformly distributed in the whole frequency domain), but different numbers of modes $K$. The adaptivity of RVMD is again proved in these trials obtained. The modes that oscillate around the fundamental screech tone are always captured. For larger $K$, more harmonics and low-frequency dynamics are successively resolved. In figure \ref{fig:param_K_init}($b$), we depict the RVMD results with different $K$, for which the central frequencies are initialized to be quadratically distributed in the frequency domain.
It is evident that compared with the result in figure \ref{fig:param_K_init}($a$), RVMD is more likely to extract the characteristic modes with relatively lower frequencies in this case, which is consistent with the prior assumption we have made in initializing the central frequencies.

According to the above discussion, we set the filtering parameter $\alpha=10000$ and the number of modes $K=24$ in the next subsection to ensure that dominant dynamic processes at various time scales are well captured. The central frequencies are initialized as uniformly distributed in the whole frequency domain, which means that no prior knowledge is required.

\subsubsection{RVMD results and discussion}
\begin{figure}
	\centerline{
		\includegraphics[width=.96\linewidth]{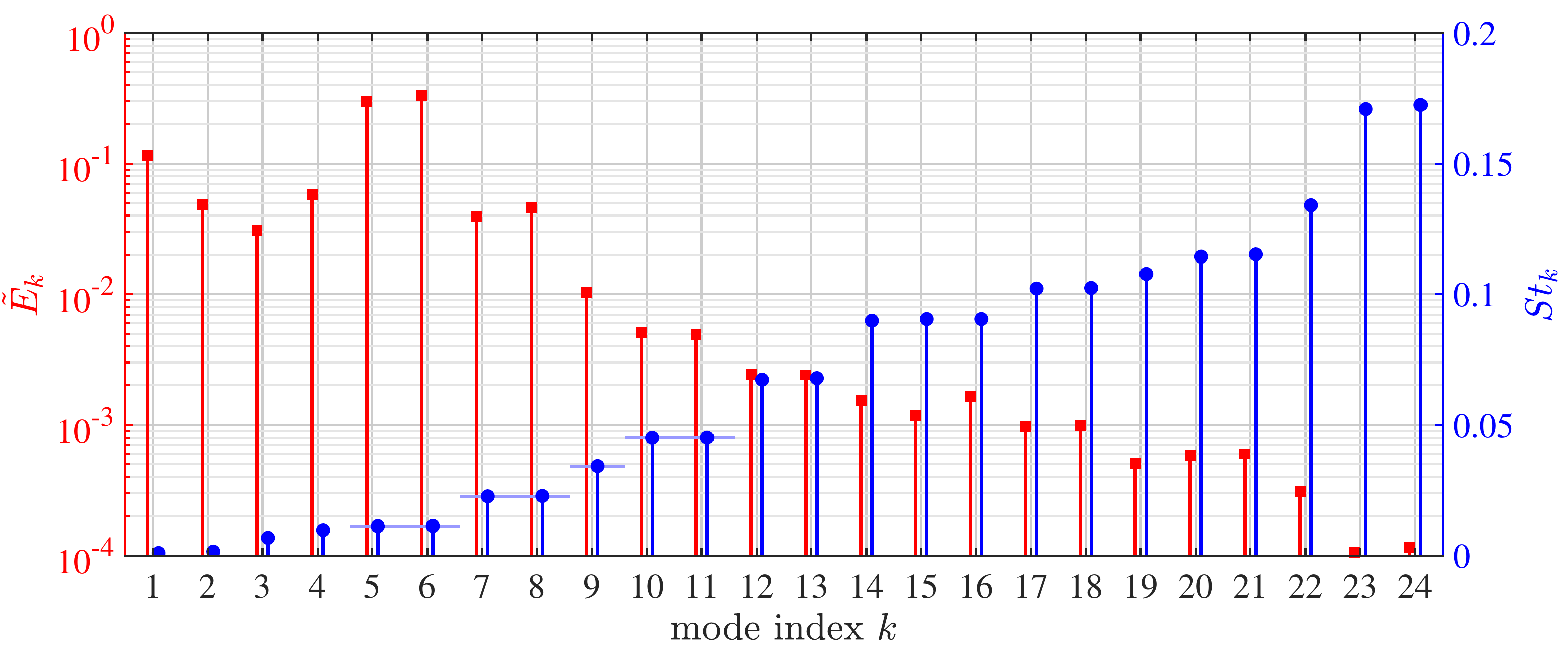}}
	\caption{RVMD results for the rectangular screeching jet: energy ratios ($\tilde{E}_k$) and central frequencies (plotted as $St_k$).}
	\label{fig:rectjetrvmd}
\end{figure}

\begin{figure}
	\centerline{
		\includegraphics[width=.96\linewidth]{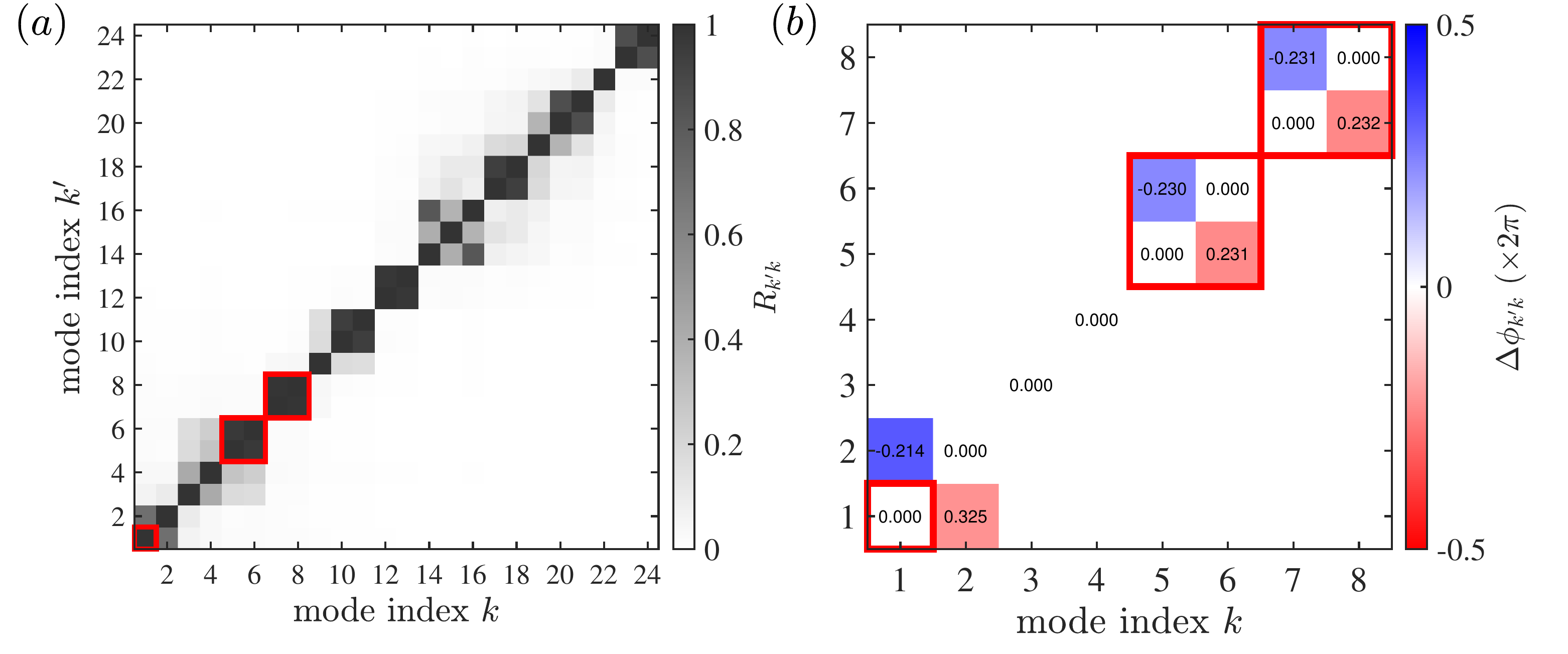}}
	\caption{($a$) The maximum cross-correlation and ($b$) the phase lag between each RVMD mode.}
	\label{fig:rectjetcorr}
\end{figure}

The energy ratios and central frequencies (namely, $St_k$) of the RVMD modes are given in figure \ref{fig:rectjetrvmd}. The fundamental screech tone ($St_s\approx0.114$) and its harmonics are also marked using horizontal light-blue line segments. As expected, the RVMD modes precisely capture the flow processes that characterize the screeching phenomena. The first two harmonics, which are also the most energetic screeching modes, are resolved by two mode pairs (modes $5,\,6$ and modes $7,\,8$). As described in the previous case, a mode pair consists of two modes for which the central frequencies and the energies contained are very close. Figure \ref{fig:rectjetcorr} shows the maximum cross-correlation and the phase lag between the RVMD modes. The maximum cross correlation between time coefficients $c_{k'}(t)$ and $c_{k}(t)$ is defined as
\begin{equation}
	R_{k'k}=\frac{\max_{\tau}\int_{-\infty}^\infty c_k(t)c_{k'}(t+\tau)\ \mathrm{d}t}{
	\sqrt{\int_{-\infty}^\infty |c_k(t)|^2\ \mathrm{d}t}\sqrt{\int_{-\infty}^\infty |c_{k'}(t)|^2\ \mathrm{d}t}
	},
\end{equation}
and the corresponding phase lag is
\begin{equation}
	\Delta\phi_{k'k}=\omega_k\tau,\quad \tau=\underset{\tau}{\operatorname{argmax}}\left\{\int_{-\infty}^\infty c_k(t)c_{k'}(t+\tau)\ \mathrm{d}t\right\}.
\end{equation}

\begin{figure}
	\centerline{
		\includegraphics[width=.96\linewidth]{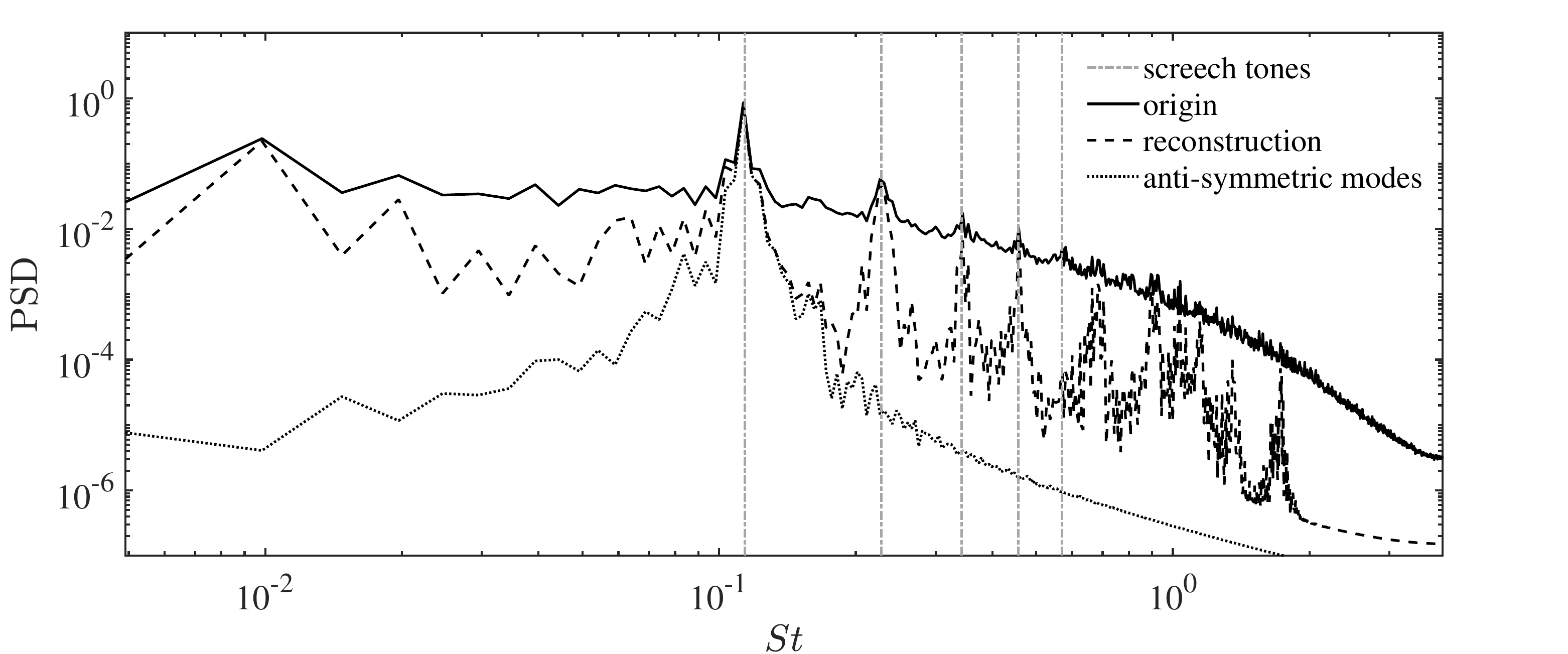}}
	\caption{Spatially averaged PSD of the rectangular screeching jet.}
	\label{fig:rectjetpsd}
\end{figure}

These two variables further quantitatively reveal the properties of the so-called mode pair. Both $R_{56}$ and $R_{78}$ exceed $0.95$, and the phase lag between the two modes in each pair is around a quarter $|\Delta\phi_{56}|\approx|\Delta\phi_{78}|\approx1/4$ like the Fourier modes $\sin\omega t$ and $\cos\omega t$. In addition to the screeching modes, some low-frequency energetic modes (modes $1,2$) are obtained, whose central frequencies are one order of magnitude lower than the fundamental screech tone. These low-frequency characteristics are also found in the power spectral density (PSD) of the pressure fluctuations close to the nozzle lip \citep{Berland07}. The spatially averaged PSD of the rectangular turbulent supersonic screeching jet and the spectrum reconstructed by the RVMD modes are depicted in figure \ref{fig:rectjetpsd}. The reconstructed PSD clearly shows the frequency-domain sparsification property of RVMD -- the distinctive dynamic properties are adaptively captured while the surrounding broadband noise (fluctuation) induced by turbulence is dropped.

\begin{figure}
	\centerline{
		\includegraphics[width=.96\linewidth]{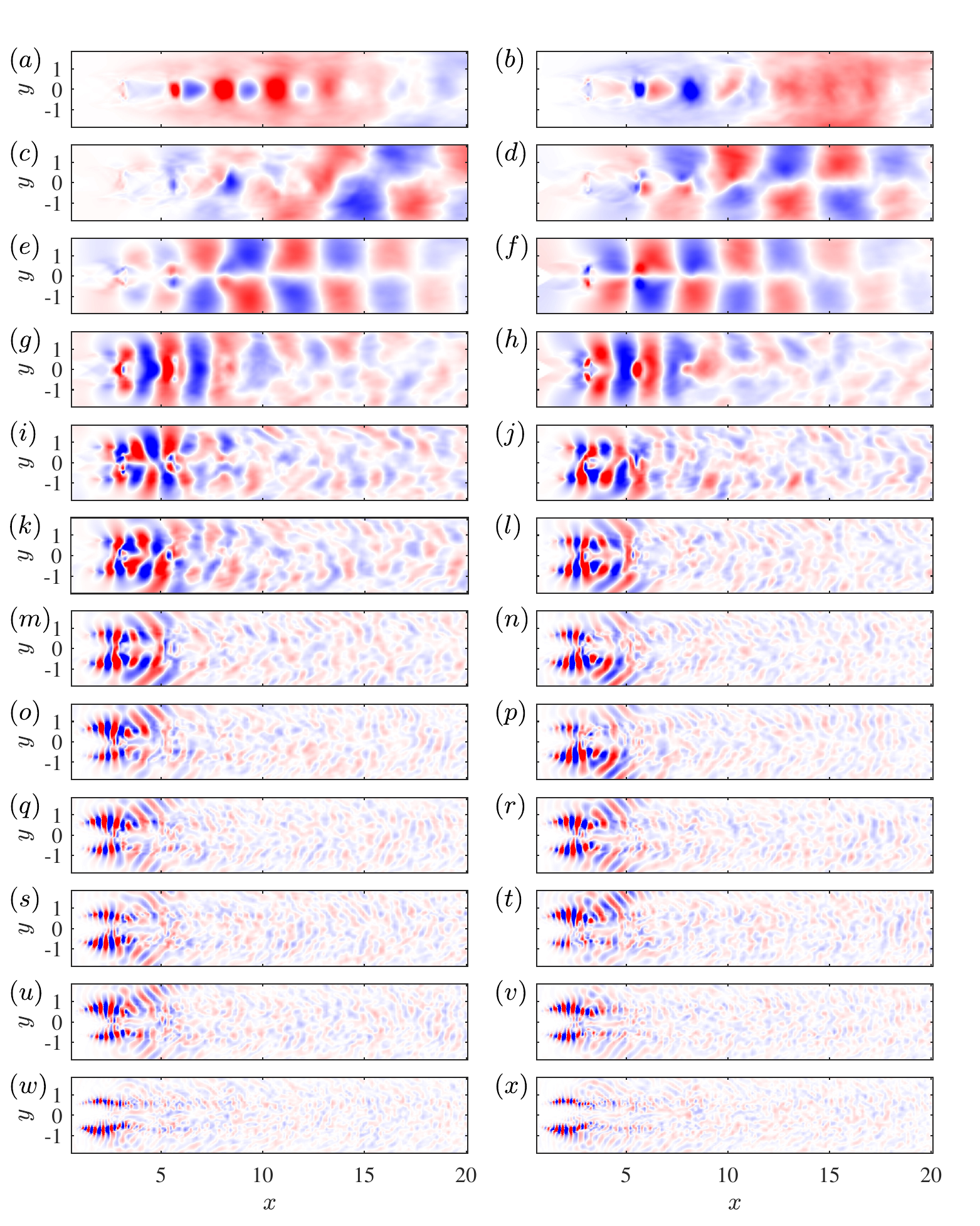}}
	\caption{Spatial distributions of RVMD modes for the rectangular screeching jet. ($a$-$x$) Modes 1-24.}
	\label{fig:rectjetphi1}
\end{figure}

Figure \ref{fig:rectjetphi1} shows the spatial distributions of the RVMD modes. The fundamental screeching modes ($5,6$) are anti-symmetric, indicating a flapping behaviour of the jet. The second-order screeching modes ($7,8$) are symmetric, indicating a nonflapping `varicose' structure. The spatially averaged PSD of the flow reconstructed by the anti-symmetric mode pair is shown in figure \ref{fig:rectjetpsd}, displaying a significant energetic dominance. These decomposed modes are in good agreement with the experimental observations in \citet{Raman94}. In Raman's work, the symmetry properties of the two screeching modes are inferred from the phase lag between pressure signals detected by a pair of microphones symmetrically arranged. Another property of the spatial distribution apart from the difference in symmetry is that the anti-symmetric modes span the entire shock-cell region, while the symmetric modes are localized around the first two/three shock cells. This spatial locality is much more pronounced for higher-order harmonic modes, as seen in figure \ref{fig:rectjetphi1}($i$-$x$). Convective wavepacket structures can be found within the shear layers on both sides of the first shock cell. When the shear layers impinge the first normal shock, the high-frequency pressure fluctuation will radiate outwards other than directly penetrating the shock to travel downstream.

\begin{figure}
	\centerline{
		\includegraphics[width=.96\linewidth]{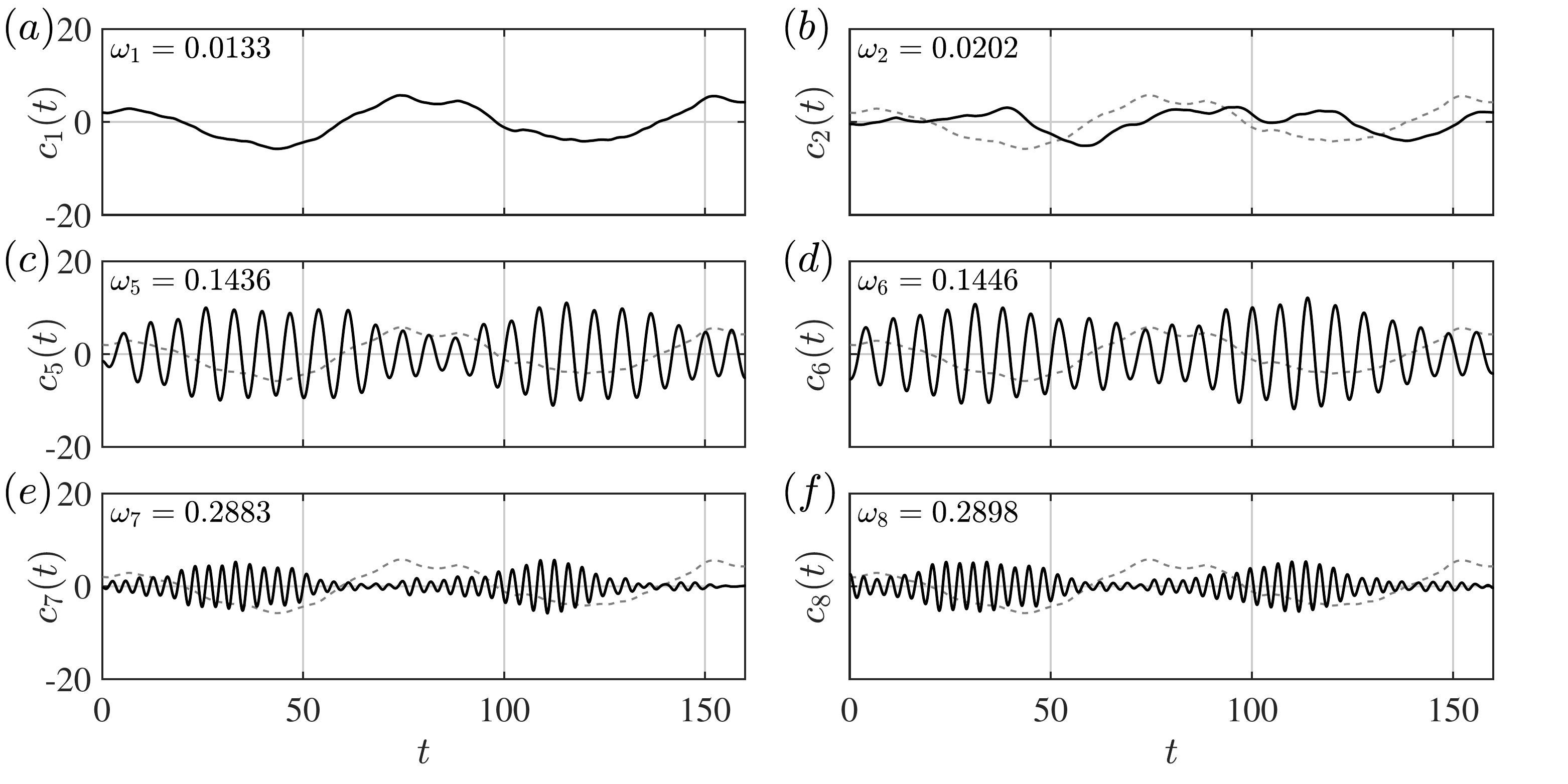}}
	\caption{Selected time-evolution coefficients of RVMD modes for the rectangular screeching jet. ($a$-$b$) Modes 1-2, ($c$-$f$) modes 5-8.}
	\label{fig:rectjetc}
\end{figure}

Interestingly, it is found in figure \ref{fig:rectjetphi1}($a,\,b$) that the two aforementioned low-frequency modes ($1,2$) actually represent an oscillatory stretching of the shock cell in the streamwise direction. The energy of these two modes is lower than the anti-symmetric modes but slightly higher than the symmetric modes (see figure \ref{fig:rectjetrvmd}), which means that this low-frequency feature should not be ignored when we consider the screeching phenomena in rectangular supersonic jets. In previous studies on screeching jets, the shock cell spacing is always treated as constant in time, forming the basis for predicting screech tones and amplitudes. Since IMFs are amplitude-modulated-frequency-modulated signals, the temporal evolution of mode strength can be obtained in RVMD modes. As depicted in figure \ref{fig:rectjetc}($c$-$f$), the envelopes of the first two screeching modes show similar oscillatory behaviour with the shock cell stretching ($a$), suggesting an intriguing interaction between the low-frequency shock-cell motion and the relatively high-frequency screeching behaviours. This amplitude modulation phenomenon can also be found in the previous simulation of \citet{Berland07}, indicated by the pressure history close to the nozzle lip. In this work, RVMD gives us an intuitive glimpse into the physical mechanism of this phenomenon. This discovery may advance the understanding and modelling of self-sustaining processes in rectangular screeching jets.

\section{Conclusion}
\label{sec:conclusion}
In this study, a novel data-driven modal analysis technique termed as reduced-order variational mode decomposition (RVMD) is proposed, which enables a low-redundant adaptive extraction of coherent structures in complex flows. The most significant advantage of this method over the existing ones is that the time-frequency information is inherently included in the RVMD modes, drawing on the concept of intrinsic mode functions in signal processing. Hence, the RVMD modes can provide a sparse representation of the observed space-time data with excellent physical interpretability.

RVMD can be regarded as the combination of the variables-separation modes and VMD, whose modes are computed by solving a minimization problem using the block coordinate descent algorithm. We prove that RVMD can be reduced to POD as the filtering parameter $\alpha$ equals zero and to DFT as $\alpha$ approaches infinity, bridging the energetic optimality and the single-frequency property. The most critical difference between RVMD and existing methods is the rich information contained in the time-evolution coefficients in IMF form. Thus, RVMD offers a succinct framework for performing time-frequency analysis on statistically nonstationary flows. A further combination of RVMD and the Hilbert spectral analysis can provide an intuitive picture of the space-time-frequency coherency of the considered flows. The properties mentioned above of RVMD have been illustrated in two canonical flow problems: the transient cylinder wake and the rectangular turbulent supersonic screeching jet. In both cases, RVMD demonstrates its adaptivity in frequency-domain sparsification. Besides, the intermediate transient processes, the linear stability modes, and the amplitude modulation phenomenon across time scales, etc., are resolved by the RVMD modes, confirming that it is a physically illuminating method.

Since RVMD is mathematically well-defined and easy to implement, further developments and applications are expected. Specifically, potential applications of this method include modal reduction of complex flows with transient behaviours, analysis of the temporal evolution and physical processes in statistically nonstationary flows, crossing-scale amplitude/frequency modulation analysis in wall turbulence, and multi-variate signal processing of fluid dynamics.

\appendix
\section{Fundamentals in signal processing}
\label{app:fundamentals}

\subsection{Wiener filtering}
\label{sec:wiener}
Here, we introduce some essential concepts in signal processing. At first, consider an observed temporal signal $f_0(t)$ consisting of the original signal $f(t)$ and a zero-mean Gaussian noise $\eta$. The inverse problem of recovering the original signal from observed data can be addressed by Tikhonov regularization, which leads to a minimization problem
\begin{equation}
	\min_f\left\{\|f-f_0\|^2+\alpha\|\partial_t f\|^2\right\},\label{eqn:wiener}
\end{equation}
where $\|\cdot\|$ denotes the $L^2$-norm, $\alpha$ is a prior parameter to control the bandwidth of the filter below. We can solve this problem through a standard variational procedure, leading to a Wiener filtering on observed data
\begin{equation}
	\hat{f}(\omega)=\frac{1}{1+\alpha \omega^2}\hat{f}_0(\omega),
	\label{eqn:filter}
\end{equation}
This simple optimization problem forms the basis of VMD and the proposed RVMD.

\subsection{Analytic signal}
\label{app:analytic}
For the convenience of mathematics, a regular manipulation is to convert the real-valued signal into a corresponding analytic representation in complex domain through Hilbert transform. The so-called analytic signal is defined as
\begin{equation}
	c_A(t)=c(t)+\mathrm{i}\mathcal{H}c(t),
\end{equation}
where $\mathrm{i}=\sqrt{-1}$ is the imaginary unit, and $\mathcal{H}$ denotes the Hilbert operator. The Hilbert transform of $c(t)$ is a linear operator defined as the Cauchy principal value (denoted by $\mathrm{p.v.}$) of the convolution with $1/\pi t$
\begin{equation}
	\mathcal{H}c(t)=\frac{1}{\pi}\mathrm{p.v.}\int_{-\infty}^\infty \frac{c(\tau)}{t-\tau}\mathrm{d}\tau.
\end{equation}
The Hilbert transform maps cosine functions to their corresponding sine functions, leading to an important property of the constructed analytic signal, unilateral spectrum (i.e., has no negative frequency components). In particular for IMF defined as (\ref{eqn:imf}), Bedrosian's theorem \citep{Bedrosian63} implies that
\begin{equation}
	c_A(t)=A(t)[\cos\varphi(t)+\mathrm{i}\sin\varphi(t)]=A(t)e^{\mathrm{i}\varphi(t)}.
\end{equation}

Another useful property of analytic signal is that multiplying it with a pure exponential results in a simple frequency shifting
\begin{equation}
	c_A(t)e^{-\mathrm{i}\omega_0t}\stackrel{\mathcal{F}}{\longleftrightarrow}\hat{c}_A(\omega)*\delta(\omega+\omega_0)=\hat{c}_A(\omega+\omega_0),
	\label{eqn:frequencyshifting}
\end{equation}
where $*$ denotes convolution, and $\delta(\cdot)$ is the Dirac delta function.
Equation (\ref{eqn:frequencyshifting}) stems from the basic modulation property of Fourier transform.

\subsection{Variational mode decomposition}
The basic idea of variational mode decomposition \citep[VMD,][]{Dragomiretskiy14} is to decompose the observed one-component real-valued signal $f(t)$ into $K$ sub-signals $u_k(t)$, namely modes, such that
\begin{enumerate}[topsep=5pt,parsep=5pt,leftmargin=20pt,listparindent=0pt,itemindent=0pt,labelsep=5pt]
	\item[(1)] The linear superposition of the $K$ modes reconstructs the input;
	\item[(2)] The bandwidth of each mode is as compact as possible. In other words, each mode is supposed to oscillate around a specific central frequency $\omega_k$.
\end{enumerate}

A constrained optimization problem can be constructed based on the two goals above and combined with the concepts outlined previously
\begin{eqnarray}
	\min_{\{u_k(t),\omega_k\}}
	\left\{
	\sum_{k=1}^K\left\|\partial_t\left\{\left[\left(\delta(t)+\mathrm{i}\frac{1}{\pi t}\right)*u_k(t)\right]e^{-\mathrm{i}\omega_k t}\right\}\right\|^2
	\right\}, \notag\\
	\mathrm{s.t.\quad}\sum_{k=1}^Ku_k(t)=f(t),
\end{eqnarray}
As the formulation in Wiener filtering (\ref{eqn:wiener}), the squared $L^2$-norm of the gradient is the bandwidth estimation of each mode. The convolution of $\delta(t)+\mathrm{i}/\pi t$ and $u_k(t)$ leads to the corresponding analytic representation of $u_k(t)$. An exponential of the carrier frequency $\omega_k$ is multiplied to shift the mode's frequency spectrum to the baseband, transforming the original low-pass filter (\ref{eqn:filter}) into a narrow-band filter around $\omega_k$.

\bibliographystyle{plainnat}
\bibliography{main}

\end{document}